\documentclass[english, prx, superscriptaddress, showpacs, reprint,]{revtex4-1}
\usepackage[latin9]{inputenc}
\usepackage{color}
\usepackage{babel}
\usepackage{verbatim}
\usepackage{bm}
\usepackage{amsmath}
\usepackage{amssymb}
\usepackage{graphicx}
\usepackage{tabularx}
\usepackage{multirow}
\usepackage{booktabs}
\usepackage{colortbl}
\usepackage{float}
\usepackage{placeins}
\usepackage{makecell}
\usepackage{enumerate}
\usepackage{amsthm}
\usepackage{amsfonts}

\definecolor{tabcolor}{rgb}{.410,.10,.11}
\usepackage[unicode=true,pdfusetitle,
 bookmarks=true,bookmarksnumbered=false,bookmarksopen=false,
 breaklinks=false,pdfborder={0 0 0},pdfborderstyle={},backref=false,colorlinks=true]
 {hyperref}
\hypersetup{
 citecolor=blue}

\usepackage{ulem}

\newcommand{\add}[1]{\textcolor{red}{{#1}}}

\makeatletter

\usepackage{epstopdf}
\usepackage{amsfonts}
\usepackage{mathrsfs}\usepackage{bm}\usepackage{appendix}\usepackage{color}
\usepackage{txfonts}
\usepackage{float}
\usepackage{epstopdf}

\makeatother
\begin{document}

\title{Chiral SO(4) spin-valley density wave and degenerate topological superconductivity in magic-angle-twisted bilayer-graphene}

\author{Chen Lu}
\affiliation{School of Physics and Technology, Wuhan University, Wuhan 430072, China}
\affiliation{School of Physics, Beijing Institute of Technology, Beijing 100081, China}
\author{Yongyou Zhang}
\email{yyzhang@bit.edu.cn}
\affiliation{School of Physics, Beijing Institute of Technology, Beijing 100081, China}
\author{Yu Zhang}
\affiliation{Shenzhen Key Laboratory of Advanced Quantum Functional Materials and Devices, Southern University of Science and Technology, Shenzhen 518055, China}
\affiliation{Department of Physics and Institute for Quantum Science and Engineering, Southern University of Science and Technology, Shenzhen 518055, China}
\author{Ming Zhang}
\affiliation{School of Physics, Beijing Institute of Technology, Beijing 100081, China}
\author{Cheng-Cheng Liu}
\affiliation{School of Physics, Beijing Institute of Technology, Beijing 100081, China}
\author{Yu Wang}
\affiliation{School of Physics and Technology, Wuhan University, Wuhan 430072, China}
\author{Zheng-Cheng Gu}
\affiliation{Department of Physics, The Chinese University of Hong Kong, Shatin, New Territories, Hong Kong, China}
\author{Wei-Qiang Chen}
\affiliation{Shenzhen Key Laboratory of Advanced Quantum Functional Materials and Devices, Southern University of Science and Technology, Shenzhen 518055, China}
\affiliation{Department of Physics and Institute for Quantum Science and Engineering, Southern University of Science and Technology, Shenzhen 518055, China}
\author{Fan Yang}
\email{yangfan\_blg@bit.edu.cn}
\affiliation{School of Physics, Beijing Institute of Technology, Beijing 100081, China}

\begin{abstract}
Starting from a realistic extended Hubbard model for a $p_{x,y}$-orbital tight-binding model on the Honeycomb lattice, we perform a thorough investigation on the possible electron instabilities in the magic-angle-twisted bilayer-graphene near the van Hove (VH) dopings. Here we focus on the interplay between the two symmetries of the system. One is the approximate SU(2)$\times$SU(2) symmetry which leads to the degeneracy between the inter-valley spin density wave (SDW) and valley density wave (VDW) as well as that between the inter-valley singlet and triplet superconductivities (SCs). The other is the $D_3$ symmetry which leads to the degeneracy among the three symmetry-related wave vectors of the density-wave (DW) orders, originating from the Fermi-surface nesting. The interplay between these two degeneracies leads to intriguing quantum states relevant to recent experiments, as revealed by our systematic random-phase-approximation based calculations followed by a succeeding mean-field energy minimization for the ground state energy. At the SU(2)$\times$SU(2) symmetric point, the degenerate inter-valley SDW and VDW are mixed into a new state of matter dubbed as the chiral SO(4) spin-valley DW. This state simultaneously hosts three 4-component vectorial spin-valley DW orders with each adopting one wave vector, and the polarization directions of the three DW orders are mutually perpendicular to one another. 
In the presence of a tiny inter-valley exchange interaction with coefficient $J_H\to 0^{-}$ which breaks the SU(2)$\times$SU(2) symmetry, a pure chiral SDW state is obtained. In the case of $J_H\to 0^{+}$, although a nematic VDW order is favored, the two SDW orders with equal amplitudes are accompanied simultaneously. This nematic VDW+SDW state possesses a stripy distribution of the charge density, consistent with the recent STM observations. On the aspect of SC, while the triplet $p+ip$ and singlet $d+id$ topological SCs are degenerate at $J_H=0$ near the VH dopings, the former (latter) is favored for $J_H\to 0^{-}$ ($J_H\to 0^{+}$). In addition, the two asymmetric doping-dependent behaviors of the obtained pairing phase diagram are well consistent with experiments.
\end{abstract}


\maketitle

\section{Introduction}
The condensed-matter community is witnessing a surge in the synthesis and research of novel graphene-multi-layer-heterostructure materials \cite{caoyuan1, caoyuan2, Caoyuan2019, Choi, Jiang, Kerelsky, Liu2019, Shen, Wangfeng1, Wangfeng2, Wangfeng3, Yankowitz, Xie} with Moir\'e pattern superstructure \cite{Bistritzer11, Lopesdos, Wangfeng4, Wu112, Xian112, new1, new2, new3, new4, new6, new7, new8, new9, new10, new12, new13, new15}, leading to the greatly enlarged unit cell and hence thousands of energy bands within the Moir\'e Brillouin zone (MBZ). Remarkably, several isolated flat bands emerge within the high-energy band gap, which brings about strong electron correlations and different types of electronic instabilities, including the correlated insulators and superconductivity (SC). Here we focus on the magic-angle-twisted bilayer-graphene (MA-TBG) \cite{caoyuan1, caoyuan2}, in which spin-unpolarized \cite{Yankowitz} correlated insulating phases are revealed when the low energy flat valence or conduction bands are half-filled, and it leads to the novel SC after doping.

Currently, the characterization of the correlated insulating phase near this doping level
\cite{bc5, bc9, bc13, caoyuan2, Fernandes, Haule234, Kang1, Kang2, PALee, Phillips, Pizarro, Schdev, Senthil, XiDai123, Xie12, Xu2, Yuan2018, Code234, Dodaro, Fanyang, Fidrysiak, Guo, Kuroki2018, Lu234, Ma, Rademaker, Honerkamp, Kennes, LiangFu, Sherkunov, add01, add02, add03, add04, add05}, the pairing mechanism, and pairing symmetry \cite{caoyuan1, Code234, Dodaro, Fanyang, Fidrysiak, Guo, Honerkamp, Kennes, Kuroki2018, LiangFu, Lu234, Ma, Rademaker, Sherkunov, Xu2018, Roy2018, Zhanglong, Ray, Lin2018, Peltonen, MacDonald, Bernevig, bc1, bc3, bc4, bc7, bc8, bc10, bc11, bc14, bc15, bc17, bc18, VishwanathYou, Stauber, Laksono, new5, new11, new2019, Chichinadze2019} are still under debate. Particularly, two opposite points of view are held, i.e. the strong-coupling Mott-insulating picture and the weak-coupling itinerant picture. Here we start from the weak-coupling viewpoint first proposed in Ref. \cite{Fanyang} that the correlated insulator and SC in the MA-TBG are driven by Fermi-surface (FS) nesting near the van Hove singularity (VHS) \cite{LiangFu, VishwanathYou, Nandkishore, Stauber, Laksono, Sherkunov, Kennes, Kozii, Yuan2019, Honerkamp, Wu2021tbg, new14, new142, new17}. The key point is that the spin or charge susceptibility would diverge as the system is doped to the VHS point with good FS-nesting, leading to the spin or charge (including valley) density wave (DW). When the doping level deviates from the DW ordered regime, the short-ranged DW fluctuations would mediate the SC. Two questions naturally arise: What type of spin or/and charge (or valley) DW would be driven by the FS-nesting near the VHS for the MA-TBG? What is the pairing symmetry mediated by the DW fluctuations?

The answers of the two questions are deeply related to the symmetries of the MA-TBG. One relevant symmetry is the $D_3$ symmetry. In the weak-coupling theories \cite{Fanyang,LiangFu, VishwanathYou}, the wave vector of the DW orders is determined by the FS-nesting vector. However, the presence of the $D_3$ symmetry brings about three degenerate FS-nesting vectors \cite{Fanyang,LiangFu,VishwanathYou}. The different DW orders hosting these degenerate wave vectors can be mixed to minimize the energy in general, leading to an exotic ground state. For example, in the theory proposed in Ref.~\cite{Fanyang}, the three SDW orders hosting degenerate wave vectors of $(0,\pi), (\pi,0)$ and $(\pi,\pi)$ would coexist and be equally mixed into the chiral SDW state, in which the polarization directions of the three vectorial SDW order parameters are mutually perpendicular and can be globally arbitrarily rotated in the $\mathbb{R}^3$ space by the Goldstone zero modes. This state breaks the time-reversal symmetry (TRS), and can be topologically nontrivial with nonzero Chern numbers.

The other relevant symmetry is associated with the special valley degree of freedom of the MA-TBG. As revealed in the continuum-theory model \cite{MacDonald}, the electron states within the two different MBZs centered at $K$ and $K^\prime$ would not hybridize for small twist angles, leading to two isolated and TR related sectors of energy bands, leading to the valley-U(1) symmetry, which survives the electron-electron interactions \cite{Yuan2018, Po2018, Kang1, LiangFu, Kuroki2018, VishwanathYou, Liang1Fu2, Po2}. Besides, this system additionally holds a spin SU(2)$_K\times$SU(2)$_{K^\prime}$ symmetry \cite{LiangFu, VishwanathYou}. Although this symmetry survives the dominant interactions in the MA-TBG, it would be slightly broken by a tiny inter-valley exchange interaction whose strength $J_H$ is much lower than any other energy scale of the system and can be treated as $J_H\to 0$. The SU(2)$_K\times$SU(2)$_{K^\prime}$ symmetry has a profound influence on the formula of the order parameters of the instabilities of the MA-TBG: it leads to the degeneracy between the inter-valley spin DW (SDW) and valley DW (VDW) as well as that between the inter-valley-pairing spin-singlet and spin-triplet SCs of the MA-TBG \cite{LiangFu,VishwanathYou}. Due to these degeneracies at the exactly-symmetric point, \sout{it's}{\color{red}it is} generally perceived that the realized instabilities in the MA-TBG are determined by the tiny $J_H$: for the case of $J_H\to 0^-$ ($J_H\to 0^+$), a pure SDW (VDW) will be the realized DW order, and a triplet $p+ip$ (singlet $d+id$) will be the pairing symmetry \cite{LiangFu,VishwanathYou}.  However, here we hold a different point of view, as introduced below.

The fact that the SDW and VDW orders are degenerate at the exactly SU(2)$_K\times$SU(2)$_{K^\prime}$-symmetric point with $J_H=0$ doesn't necessitate that only one of them is the candidate for a tiny $J_H$. Actually, the two orders can generally be mixed to lower the ground-state energy in any case. The right procedure for the identification of the ground-state DW orders for different $J_H$ is as follow. Firstly, we should identify the energetically minimized mixing manner between the SDW and VDW at the symmetric point with $J_H=0$. Note that the mixing manner thus obtained is not unique, as the spontaneous breaking of the SU(2)$_K\times$SU(2)$_{K^\prime}$ symmetry always leads to gapless Goldstone modes which can rotate one ground state to numerous other degenerate ones, forming a ground-state subspace. Then the realistic tiny $J_H$-term sets in, which serves as a perturbative symmetry-breaking field and will select its favorite states from this subspace. These states form the ground states for nonzero $J_H$. Note that the $D_3$ symmetry plays an important role in this procedure: it will introduce three times as many states to participate in the mixing, which fundamentally changes the ground state. The ground state thus obtained turns out to be fundamentally different from the intuitively conjectured one in Ref.~\cite{LiangFu,VishwanathYou}.

In this paper, we perform a thorough investigation on the DW orders and SC in the MA-TBG driven by FS-nesting near the VHS, with a particular attention paid to the interplay between the approximate SU(2)$_K\times$SU(2)$_{K^\prime}$ symmetry and the threefold degeneracy among the wave vectors of the DW orders. Through adopting realistic band structure and interaction terms that respect all symmetries of the system, we carry out systematic calculations based on the random-phase approximation (RPA) and subsequent mean-field (MF) energy minimization for the ground state. We have also provided a phenomenological Ginzburg-Landau theory to account for our microscopic results. While the RPA calculations suggest that the critical interactions $U^{(s)}_c$ and $U^{(v)}_c$ for the SDW and VDW orders are equal at $J_H=0$, the subsequent MF energy minimization yields that the SDW ground state holds a lower energy because its vectorial order parameters allow three times as many states to participate in the mixing and thus have more opportunity to lower the energy. When we further allow the SDW and VDW to mix, a novel chiral SO(4) spin-valley DW state with exotic properties is obtained, as will be introduced in Sec. II. When the tiny inter-valley exchange interaction term is added, we obtain the pure chiral SDW state for $J_H\to 0^-$ and a nematic DW state with mixed SDW and stripy VDW orders for $J_H\to 0^+$. The latter case is consistent with the recent STM experiment \cite{Jiang,Kerelsky}, and might be more probably realized in the MA-TBG. On the $J_H$-dependent pairing symmetries, our results are essentially consistent with the intuitively conjectured one in Ref.~\cite{LiangFu,VishwanathYou}.

The rest of this paper is organized as follows. Section II provides an overview on the main results provided in this work. In Sec. III, we describe the model and the approach. A two-orbital tight-binding (TB) model on the honeycomb lattice is provided, added with realistic interaction terms. The RPA approach and the subsequent MF analysis are introduced. In Sec. IV, we study the case of $J_H=0$, in which the system hosts the exact SU(2)$_K\times$SU(2)$_{K^\prime}$ symmetry. The degeneracies between the SDW and VDW as well as between the singlet and triplet SCs are analyzed in detail. We find that the SDW and VDW can mix into the chiral SO(4) spin-valley DW. In Sec. V, we provide our results for the cases with tiny $J_H\ne 0$, including $J_H\to 0^{+}$ and $J_H\to 0^{-}$. These two cases have different DW states and pairing symmetries. Finally, a conclusion will be reached with some discussions in Sec. VI.

\begin{figure}
\centering
\includegraphics[width=0.48\textwidth]{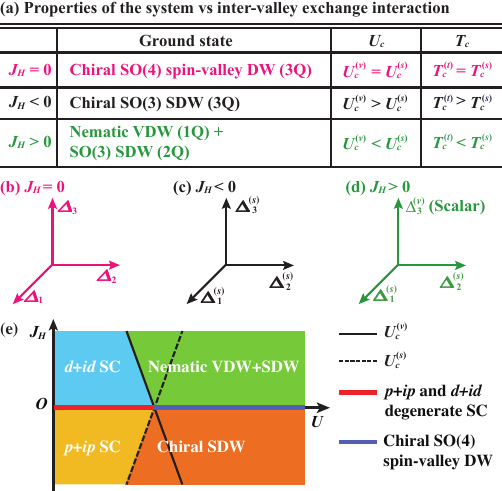}
\caption{(a) The properties of the system, including the characterization of the ground state, the relation between the critical interactions $U_c$ for VDW $\left(U_c^{(v)}\right)$ and SDW $\left(U_c^{(s)}\right)$ as well as that between the $T_c$ of singlet-$\left(T_c^{(s)}\right)$ and triplet-$\left(T_c^{(t)}\right)$ SCs, for different inter-valley exchange interactions. (b-d) The corresponding DW order-parameter configurations of the ground states. In panel (a) the number after SDW and VDW denotes how many $\bm Q_\alpha$ are distributed to the corresponding DW orders. When $J_H=0$ the ground state is in the chiral SO(4) spin-valley DW phase, wherein the three mutually perpendicular four-dimensional order-parameter vectors $\bm{\varDelta}_{\alpha}=\left(\Delta^{(v)}_{\alpha},  \Delta^{(s)}_{\alpha,x},\Delta^{(s)}_{\alpha,y},\Delta^{(s)}_{\alpha,z}\right)$ can be globally arbitrarily rotated in the $\mathbb{R}^4$ VDW-SDW order-parameter space by the Goldstone zero modes, see panel (b). When $J_H<0$ the ground state is in the chiral SDW phase, wherein the three mutually perpendicular SDW vectors $\bm \Delta_\alpha^{(s)}=\left(\Delta^{(s)}_{\alpha,x},\Delta^{(s)}_{\alpha,y},\Delta^{(s)}_{\alpha,z}\right)$ can be globally arbitrarily rotated in the $\mathbb{R}^3$ SDW space, see panel (c). When $J_H>0$, one wave vector, e.g. $\bm{Q}_3$, is fully occupied by the scalar VDW order $\Delta_3^{(v)}$, and the remaining two are occupied by the vectorial SDW orders, i.e. $\bm \Delta_1^{(s)}$ and $\bm \Delta_2^{(s)}$, which are perpendicular to each other and can be globally arbitrarily rotated in the $\mathbb{R}^3$ SDW order-parameter space, see panel (d). The schematic phase diagram with respect to the $U$-$J_H$ parameters are shown in (e).}
\label{classification}
\end{figure}

\section{Overview}

This section provides an overview on the present work, which is focused on how the interplay between the approximate SU(2)$_K\times$SU(2)$_{K^\prime}$ symmetry and the $D_3$ symmetry will influence the formula of the order parameters of the DW and SC in the MA-TBG. Briefly speaking, our answer to the question about the DW is fundamentally different from the generally perceived one. Due to the degeneracy between the SDW and VDW and that between singlet and triplet pairings at the exact SU(2)$_K\times$SU(2)$_{K^\prime}$-symmetric point with $J_H=0$, \sout{it's}{\color{red}it is} generally intuitively perceived that for the case of $J_H\to 0^-$ ($J_H\to 0^+$), a pure SDW (VDW) will be realized, and a triplet $p+ip$ (singlet $d+id$) will be the pairing symmetry \cite{LiangFu,VishwanathYou}.  However, here we propose that the two DW orders are generally mixed. In the case of $J_H=0$, we obtained the chiral SO(4) spin-valley DW, which evolves into a pure chiral SDW upon $J_H\to 0^-$ and a nematic DW with mixed SDW and stripy VDW orders upon $J_H\to 0^+$. The latter case is consistent with recent STM observations. For the SC, our answer is consistent with the generally perceived viewpoint.

Our start point is the $p_{x,y}$-orbital tight-binding (TB) model on the Honeycomb lattice \cite{Yuan2018, Liang1Fu2}, equipped with realistic extended Hubbard interactions including a tiny inter-valley exchange interaction. While the TB part and the dominant interactions in this Hamiltonian possess the SU(2)$_K\times$SU(2)$_{K^\prime}$ symmetry, which is broken by the tiny inter-valley exchange interaction. Besides, the model holds a $D_3$ symmetry, which leads to three degenerate FS-nesting vectors $\bm Q_\alpha \ (\alpha=1,2,3)$ near the VHS points. In our calculations, we first carry out systematic RPA based studies to figure out the forms of all possible instabilities, and then perform a subsequent MF energy minimizations to pin down the mixing manner between degenerate orders. Finally, in order to account for the results obtained by our microscopic calculations, we have also provided a phenomenological Ginzburg-Landau theory to classify all the possible configurations of the DW order parameters, which emerge as possible solutions to minimize the G-L free energy function. Our results are summarized in Fig.~\ref{classification}.

The results for the case of $J_H=0$ are listed in the row of $J_H=0$ in Fig.~\ref{classification}(a). In this case, the critical interactions $U^{(s)}_c$ and $U^{(v)}_c$ for the SDW and VDW orders are equal, and the leading spin-singlet ($d+id$) and spin-triplet ($p+ip$) pairings have equal $T_c$. The degeneracy between the SDW and VDW makes them mix into the chiral SO(4) spin-valley DW ordered state. This DW state is characterized by three coexisting four-component vectorial order parameters $\bm \varDelta_\alpha$ ($\alpha=1,\ 2,\ 3$) shown in Fig.~\ref{classification}(b), with each $\bm \varDelta_\alpha\equiv \left(\Delta^{(v)}_\alpha, \Delta^{(s)}_{\alpha, x}, \Delta^{(s)}_{\alpha, y}, \Delta^{(s)}_{\alpha, z}\right)$ hosting one wave vector $\bm Q_\alpha$. Here, $\Delta^{(s)}_{\alpha, x/y/z}$ and $\Delta^{(v)}_{\alpha}$ represent the SDW and VDW order parameters hosting the wave vector $\bm Q_\alpha$, respectively. The three 4-component vectorial order parameters are mutually perpendicular to one another, i.e. $\bm \varDelta_1\perp \bm \varDelta_2 \perp \bm \varDelta_3$, and can be globally arbitrarily rotated in the $\mathbb{R}^4$ order-parameter space by the Goldstone zero modes, as shown in Fig.~\ref{classification}(b). This phase is a generalization of the 3Q chiral SDW state proposed previously \cite{Fanyang,TaoLi,Martin,Kato,Ying} to the $\mathbb{R}^4$ VDW-SDW order-parameter space, and represents a new state of matter that possesses a series of intriguing properties. For example, this DW ground state hosts seven branches of gapless Goldstone modes. In addition, the topological properties of this DW state can be nontrivial with nonzero Chern number, as long as a DW gap opens at the Fermi level.

The results for $J_H\to 0^{-}$ (Hund-like) are listed in the row of $J_H<0$ in Fig.~\ref{classification}(a). In this case, our RPA calculation yields $U^{(v)}_c>U^{(s)}_c$, suggesting that the SDW is preferred to the VDW. Therefore, in the $\mathbb{R}^4$ VDW-SDW order-parameter space, the VDW axis becomes the ``difficult'' axis and would be kicked out from the low-energy degree of freedom. As a result, our subsequent MF energy minimization yields the pure 3Q chiral SDW state characterized as $\bm \varDelta_\alpha=\left(0,\bm \Delta_\alpha^{(s)}\right)\equiv\left(0, \Delta^{(s)}_{\alpha, x}, \Delta^{(s)}_{\alpha, y}, \Delta^{(s)}_{\alpha, z}\right)$, with $\bm \Delta_1^{(s)}\perp \bm \Delta_2^{(s)} \perp \bm \Delta_3^{(s)}$, as shown in Fig.~\ref{classification}(c).  This state is qualitatively the same as that obtained previously \cite{Fanyang,TaoLi,Martin,Kato,Ying}, which hosts four branches of gapless Goldstone modes one. The Chern number can also be nonzero, as long as an SDW gap opens at the Fermi level. As for the SC, the triplet SC with $p+ip$ pairing symmetry is preferred.

The results for $J_H\to 0^{+}$ (anti-Hund-like) are listed in the row of $J_H>0$ in Fig.~\ref{classification}(a). In this case, our RPA calculation yields $U^{(v)}_c<U^{(s)}_c$, suggesting that the VDW is preferred to the SDW. Therefore, in the $\mathbb{R}^4$ VDW-SDW order-parameter space, the VDW axis becomes the ``easy'' axis. However, this doesn't suggest a pure VDW state as generally perceived \cite{LiangFu,VishwanathYou}, because here we have three 4-component vectorial DW order parameters, which can not all point along the ``easy" VDW axis, as their mutual perpendicular relation is robust against the tiny $J_H$ term. Our subsequent MF energy minimization yields a DW state with one scalar VDW component mixed with two mutually perpendicular vectorial SDW components with equal amplitude, with the VDW randomly choosing one wave vector $\bm{Q}_\alpha$ from the three symmetry-related ones and the two SDW hosting the remaining two. Obviously, this nematic DW state spontaneously breaks the $C_3$ rotation symmetry, and the obtained stripy charge order is consistent with recent STM experiments \cite{Jiang, Kerelsky}. This DW state is schematically shown in Fig.~\ref{classification}(d). The number of Goldstone modes and the topological properties in this case are the same as those in $J_H\to 0^{-}$. As for the SC, the singlet SC with $d+id$ pairing symmetry is preferred.

The schematic phase diagram with respect to the $U$-$J_H$ parameters are shown in Fig.~\ref{classification}(e). Besides the $J_H$-dependence, our results reveal two asymmetric doping-dependent behaviors in the pairing phase diagram. One is the asymmetry with respect to the charge neutral point (CNP): the $T_c$ at the negative dopings is much higher than that at the positive dopings, which is due to the higher DOS in the former case. The other asymmetry is with respect to each VH doping: the $T_c$ on the higher-doping side of each VH point is higher than that on its lower-doping side. This asymmetry is attributed to the better FS-nesting and hence stronger DW fluctuations in the former case. These two asymmetric doping-dependent behaviors are well consistent with the experiments \cite{caoyuan1,Yankowitz}, implying that the pairing in the MA-TBG should be mediated by the spin-valley DW fluctuations.

\section{Model and Approach}
\begin{figure}
\centering
\includegraphics[width=0.47\textwidth]{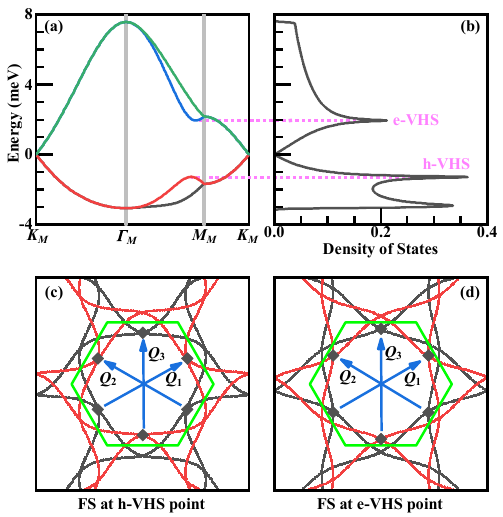}
\caption{Band structure of the TB model \eqref{HTB} representing the MA-TBG. (a) The band structure along the high-symmetry lines, with the CNP setting as the zero point of energy. (b) The corresponding DOS, with the two VHS points denoted as h-VHS and e-VHS representing for the VHSs of the hole- and electron- dopings, respectively. (c, d) FSs at the h-VHS and e-VHS doping levels with $\delta=$ -0.182 and 0.240, respectively. The green hexagon represents the MBZ. The black and red curves correspond to the FSs from the $K$ and $K'$ valleys, respectively. The three $\bm Q_\alpha$ in blue mark the FS-nesting vectors. The TB parameters adopted are $t_1=1.5$ meV, $t'_1=-0.8$ meV, $t_2=0.25$ meV, $t'_2=0$, $t_3=0.2$ meV, and $t'_3=0.3$ meV.}\label{band}
\end{figure}

\subsection{Model}\label{secModel}

For the MA-TBG there are four low-energy flat bands that are well isolated from the high-energy bands\cite{Nguyen2017, Moon2012, Fang2015, Lopesdos, Santos2012, Shallcross2008, Bistritzer11, Bistritzer2010, Uchida2014, Mele2011, Mele2010, Sboychakov2015, Morell2010, Trambly2010, Latil2007, Trambly2012, Huang123, Guinea234, Gonza2013, Luis2017, Cao2016, Ohta2012, Kim2017, Huder2018, Li2017, Senthil, Yuan2018, Bernevig2018, Balents2018, Po2018, Zhanglong, Ray, VishwanathYou, Kang1, Liang1Fu2, Pal2018, Guinea2018, Zou2018, Po2, Tarnopolsky, Ahn, Morell2010, Trambly2012, bc12}. The four flat bands can be divided into two valence bands and two conduction bands, which touch at the charge neutral point (CNP), i.e., $K_M$ and $K_M^\prime$ points in the MBZ. Besides the four-fold degeneracy at the CNP, the valence and conduction bands each are two-fold degenerate along the $\varGamma_MK_M$ and $K_MM_M$ lines. The continuum theory \cite{Bistritzer11, Bistritzer2010} tells that these degeneracies are the consequence of the so-called U(1)-valley symmetry of the TBG with small twist angles. This symmetry forbids the hopping from the MBZ in the $K$ valley to that in the $K'$ valley. While the TB models in Ref. \cite{Po2} can faithfully describe the low-energy flat bands in both aspects of the symmetry and the topology at the CNPs, they are too complicated to be sufficiently convenient for succeeding studies with electron-electron interactions. Here we focus on the low-energy band structure near the Fermi level for the doped case, particularly near the VHS points which are related to experiments, which allows us to adopt simpler band structures.

The proposed simplest TB model for the MA-TBG is that on the honeycomb lattice containing a $p_x$- and a $p_y$-orbitals on each site \cite{Yuan2018, Po2018, Kang1, Fanyang, Liang1Fu2}, with the orbitals on adjacent cites coupling via coexisting $\sigma$- and $\pi$- bondings \cite{Fanyang}. It's proved in Appendix \ref{appendixHTB} that the valley-U(1) symmetry requires that the amplitudes of the $\sigma$- and $\pi$- bondings are equal. In such a condition,  we transform the $p_{x,y}$-representation into the valley representation by $\hat c_{j\pm\sigma}=(\hat c_{jx\sigma}\pm i\hat c_{jy\sigma})/\sqrt{2}$, where $\hat c_{j\mu\sigma}$ is the annihilation operator of the electron on the $j$-th site with spin $\sigma$ and orbital $\mu$ ($\mu=x,y$ represents the $p_x$ or $p_y$ orbital) and $\pm$ represent the $K$ and $K'$ valleys. Consequently, we can find the following TB Hamiltonian \cite{Yuan2018,Liang1Fu2},
\begin{align}\label{HTB}
\hat H_{\rm TB}
&=\sum_{\alpha}\sum _{\langle jj'\rangle_\alpha\nu\sigma}\left[\left(t_\alpha {-}i\nu t'_\alpha\right)\hat c^{\dagger}_{j\nu\sigma}\hat c_{j'\nu\sigma}{+} {\rm h.c.}\right]-\mu_c\sum _{jv\sigma}\hat c^{\dagger}_{jv\sigma}\hat c_{jv\sigma},\nonumber\\
&=\sum _{mv\bm k\sigma}\tilde\varepsilon^{mv}_{\bm k}\hat c^{\dagger}_{mv\bm k\sigma}\hat c_{mv\bm k\sigma}.
\end{align}
More details are provided in Appendix \ref{appendixHTB}. Here, $\hat c_{mv\bm k\sigma}$ is the annihilation operator of the electron with the band index $m$, the valley index $v$,  the wave vector $\bm k$ and the spin $\sigma$. The energy $\tilde\varepsilon^{mv}_{\bm k}$ is with respect to the chemical potential $\mu_c$. $\langle jj'\rangle_\alpha$ denotes the $\alpha$-th neighboring bond. $t_\alpha$ is the hopping strength that is caused by the $\sigma$ and $\pi$ bonding \cite{DMSarma, CMWu,GMF, CMCMLiu, FMYang} and $t'_\alpha$ is responsible for the Kane-Mele type of the valley-orbital coupling \cite{Yuan2018,Liang1Fu2}. In our calculations, we consider up to the third-neighbor hoppings, i.e. $\alpha=1,2,3$. The chemical potential $\mu_c$ is determined by the doping $\delta\equiv n/n_s-1$ with respect to the CNP. $n$ is the average electron number per unit cell with $n=n_s\equiv4$ for the CNP.

The TB model in Eq.~\eqref{HTB} tells that the $K$ and $K'$ valley bands are separated with each other, leading to a valley-U(1) symmetry. Moreover, each valley independently supports the spin-SU(2) symmetry, leading to an SU(2)$_K\times$SU(2)$_{K^\prime}$ symmetry. Finally, the geometry of the TBG leads to a $D_3$ point group. Figure \ref{band}(a) shows the corresponding band structure with the TB parameters provided in the figure caption. As a result of the U(1)-valley symmetry, $K_M$ points are four-fold degenerate, and $\varGamma_M$ and $M_M$ points are doubly degenerate. The U(1)-valley symmetry is also responsible for the double degeneration of the $\varGamma_MK_M$ and $K_MM_M$ lines. These characters are consistent with the continuum theory. The hump and dip in the two middle bands along the $\varGamma_MM_M$ line give two VHS points for the hole and electron dopings respectively, see Fig.~\ref{band}(b). They, denoted as the h-VHS and e-VHS, are both near the $M_M$ points and correspond to the doping of -0.182 and 0.240, respectively. These two VHSs originate from the the Lifshitz transition points, which can be seen from the FSs in Figs.~\ref{band}(c) and \ref{band}(d). The valley-separated FSs reflect the inter-valley nesting behavior whose three nesting vectors are marked as $\bm Q_\alpha$ ($\alpha =1,\ 2,\ 3$). These nesting vectors do not exactly connect the $M_M$ points, different from the previous model in Ref.~\cite{Fanyang}.

Note that in the Supplementary Material \cite{SM}, we provide the FSs at the e-VH and h-VH dopings for five different twist angles near the magic angle, which is 1$^\circ$, 1.05$^\circ$, 1.1$^\circ$, 1.15$^\circ$ and 1.2$^\circ$. The band structure is obtained via the continuum model\cite{Bistritzer11}. The resulting FSs clearly exhibit the presence of the Lifshitz transitions, which leads to the VHSs. What's more, in these FSs there are also approximate FS nesting with three-folded rotation symmetry related nesting vectors $\bm Q_\alpha$($\alpha =1,\ 2,\ 3$) whose exact values depend on the twist angles.

Symmetry analysis and the extended character of the Wannier bases \cite{Po2018, Liang1Fu2,Kuroki2018} suggest the following interaction terms for the MA-TBG,
\begin{align}\label{interaction}
\hat H_{\rm int}=& U\sum _{jv}\hat n_{jv\uparrow}\hat n_{jv\downarrow} + V\sum _j\hat n_{j+} \hat n_{j-} + \sum_{\alpha=1}^3 W_\alpha\sum _{ \langle jj' \rangle_\alpha}\hat n_j\hat n_{j'}   \nonumber\\
&{-} J\sum_{\langle jj'\rangle_1} \sum_{vv' \sigma\sigma'}\hat c^{\dagger}_{jv\sigma}\hat c_{jv'\sigma'}\hat c^{\dagger}_{j'v'\sigma'}\hat c_{j'v\sigma}    \nonumber\\
&{-} J_H\sum_{jv\sigma\sigma'}\hat c_{jv\sigma}^\dag\hat c_{j\bar v\sigma}\hat c_{j\bar v\sigma'}^\dag\hat c_{jv\sigma'},
\end{align}
where $\hat n_j=\hat n_{j+}+\hat n_{j-}$, $\hat n_{jv}=\hat n_{jv\uparrow}+\hat n_{jv\downarrow} $, and $\hat n_{jv\sigma}=\hat c^\dag_{jv\sigma}\hat c_{jv\sigma}$. The extended density-density interactions between neighboring sites are represented by the $W_\alpha$ terms which are up to the third neighbor. The relation among $W_{\alpha}$ and $U$ is assumed to be $U:W_1:W_2:W_3=3:2:1:1$ \cite{Kuroki2018, Liang1Fu2}. The exchange interaction $J=0.2U$ is taken according to Ref.~\cite{Liang1Fu2}. The tiny inter-valley Hund's-rule exchange interaction is given by the last term with the coefficient $J_H$ two orders of magnitude weaker than $U$ \cite{JYLee}, and the parameters $U$, $V$ and $J_H$ satisfy the relation $U=V+2J_H$.

The model \eqref{interaction} provides a realistic description for the electron-electron interactions in the MA-TBG. The total Hamiltonian of the system is given by
\begin{align}\label{H_Hubbard}
\hat H = \hat H_{\rm TB} + \hat H_{\rm int}.
\end{align}
Note that all the terms except the tiny $J_H$ term conserve the SU(2)$_K\times$SU(2)$_{K^\prime}$ symmetry, which is broken by the tiny $J_H$ term to the valley-U(1) symmetry plus the global spin-SU(2) symmetry. In our study, we considered the three different cases, i.e. $J_H=0$, $J_H=0.01U$ and $J_H=-0.01U$, for comparison. As will be seen below, the three different cases will lead to qualitatively different ground states. In realistic material, the interaction strength $U$ is estimated to be comparable with the band width\cite{caoyuan1}. Although in some study\cite{Kang2} the $U$ is estimated to be about an order of magnitude larger than that adopted here, the band width of the MA-TBG measured by the STM\cite{Xie} is also an order of magnitude larger than that adopted here. The experimentally-measured bandwidth can be viewed as that renormalizd by electron-electron interaction, and our TB band structure can also be viewed as the one renormalized by interaction.  Therefore, our model can be viewed as rescaled from the realistic material by a factor of about 10. Such rescaling will not alter the qualitative behavior of the system

\subsection{The RPA$+$MF approach}

The RPA approach is used in this work to identify the electron instabilities driven by the FS-nesting and VHS. According to the standard multi-orbital RPA approach \cite{RPA1, RPA2, RPA3, Kuroki, Scalapino1, Scalapino2, Liu2013, Wu2014, Ma2014, Zhang2015}, the following bare susceptibility is defined for the non-interacting case, namely,
\begin{align}\label{chi0}
\chi^{(0)l_1 l_2}_{l_3l_4}(\bm{q},\tau)\equiv
&\frac{1}{N}\sum_{\bm{k}_1\bm{k}_2}\left\langle
T_{\tau}\hat c_{l_1\bm k_1\sigma}^{\dagger}(\tau)
\hat c_{l_2\bm k_1+\bm q\sigma}(\tau)\right.             \nonumber\\
&\qquad\qquad\left.\times \hat c_{l_4\bm k_2+\bm q\sigma}^{\dagger}(0)
\hat c_{l_3\bm k_2\sigma}(0)\right\rangle_0,
\end{align}
where $\bm q$ and $\bm k_{1,2}$ are the wave vectors and $l_{1,...,4}=(\iota v)$ with $\iota=$ A and B representing the sublattice index and $v=\pm$ denoting the $K$ and $K'$ valleys respectively. The $\langle\cdots \rangle_0$ denotes the thermal average of the noninteracting system. The explicit formula of $\chi^{(0)l_1 l_2}_{l_3l_4}(\bm{q},\tau)$ is given in the Appendix \ref{RPAapproach}.

When interactions turn on, we define the following renormalized spin and charge susceptibilities,
\begin{subequations}\label{SUS}
\begin{align}
\chi^{(s)l_1l_2}_{l_3l_4}\left(\bm{q},\tau\right)
&\equiv{1\over 2N} \sum_{\bm{k}_1\bm{k}_2,\sigma_1\sigma_2} \left\langle
T_{\tau}\hat c_{l_1\bm{k}_1\sigma_1}^{\dagger}(\tau)
\hat c_{l_2\bm{k}_1+\bm{q}\sigma_1}(\tau)\right.\nonumber\\
&\qquad\left.\times \hat c_{l_4\bm{k}_2+\bm{q}\sigma_2}^{\dagger}(0)
\hat c_{l_3\bm{k}_2\sigma_2}(0)\right\rangle\sigma_1\sigma_2 ,\label{chis}\\
\chi^{(c)l_1l_2}_{l_3l_4}\left(\bm{q},\tau \right)
&\equiv\frac{1}{2N}\sum_{\bm{k}_1\bm{k}_2,\sigma_1\sigma_2}\left\langle
T_{\tau}\hat c_{l_1\bm{k}_1\sigma_1}^{\dagger}(\tau)
\hat c_{l_2\bm{k}_1+\bm{q}\sigma_1}(\tau)\right.\nonumber\\
&\qquad\left.\times \hat c_{l_4\bm{k}_2+\bm{q}\sigma_2}^{\dagger}(0)
\hat c_{l_3\bm{k}_2\sigma_2}(0)\right\rangle  .\label{chic}
\end{align}
\end{subequations}
In the RPA level, they are related to the bare susceptibility through the relation
\begin{subequations}
\begin{align}
\chi^{(s)}\left(\bm{q},i\omega \right)=&\left[I-\chi^{(0)}
\left(\bm{q},i\omega \right)\tilde{U}^{(s)}\right]^{-1}\chi^{(0)}
\left(\bm{q},i\omega \right),\label{Renorm_SUS_spin}\\
\chi^{(c)}\left(\bm{q},i\omega \right)=&\left[I+\chi^{(0)}
\left(\bm{q},i\omega \right)\tilde{U}^{(c)}\right]^{-1}\chi^{(0)}
\left(\bm{q},i\omega \right).\label{Renorm_SUS_charge}
\end{align}
\end{subequations}
Here, $\chi^{(0)/(s)/(c)}(\bm{q},i\omega)$ are the Fourier transformations of $\chi^{(0)/(s)/(c)}(\bm{q},\tau)$ in the imaginary-frequency space, which are operated as $16\times 16$ matrices by taking the upper and lower two indices as one number, respectively. Note that we only provide the $zz$-component of the spin susceptibility. In the presence of spin-SU(2) symmetry, the other two components, i.e. the $+-$ and $-+$ components are equal to the $zz$ component. The forms for $\tilde{U}^{(s)/(c)}$ are given in Appendix \ref{RPAapproach}.

If $U>U^{(s)}_{c}$ $\left(U>U^{(c)}_{c}\right)$, the denominator matrix in Eq.~\eqref{Renorm_SUS_spin} (Eq.~\eqref{Renorm_SUS_charge}) has zero eigenvalue(s) for some $(\bm{q},i\omega=0)$ and the renormalized zero-frequency spin (charge) susceptibility $\chi^{(s)}$ $\left(\chi^{(c)}\right)$ diverges, implying the formation of DW order in the spin (charge) channel. The concrete formulism of the interaction-induced DW order in the spin (charge) channel can be constructed as follow.

Let $U\to U^{(s)}_{c}$ ($U\to U^{(c)}_{c}$) from below, get the eigenvector $\xi^{(s)}(\bm{Q})$ $\left(\xi^{(c)}(\bm{Q})\right)$ corresponding to the largest eigenvalue of $\chi^{(s)}(\bm{Q}, i\omega=0)$ $\left(\chi^{(c)}(\bm{Q}, i\omega=0)\right)$. Here the momentum $\bm{Q}$, at which $\chi^{(s)}(\bm{Q}, i\omega=0)$ $\left(\chi^{(c)}(\bm{Q}, i\omega=0)\right)$ first diverges, provides the wave vector of the interaction-induced  magnetic (valley) order, and the eigenvector $\xi^{(s)}(\bm{Q})$ ($\xi^{(c)}(\bm{Q})$) provides the form factor of the induced order. Generally in the weak-coupling limit, the wave vector $\bm{Q}$ of the interaction-induced order is equal to the FS-nesting vector. Due to the three-folded rotational symmetry of the system, there exist three degenerate FS-nesting vectors $\bm Q_\alpha$ with $\alpha =1,\ 2,\ 3$, and so do the wave vectors of the induced order. As a result, the interaction-induced SDW or CDW order can be described by the following order-parameter part of the Hamiltonian,
\begin{align}\label{HDW}
\hat H_{\rm CDW}&=\sum_{\alpha=1}^3\!\sum_{l_1l_2\bm{k} \sigma}\Delta_\alpha^{(c)}
\hat c^{\dagger}_{l_1\bm{k}\sigma}
\xi^{(c)}_{l_1l_2}(\bm{Q}_\alpha)
\hat c_{l_2\bm{k}-\bm{Q}_\alpha \sigma}+{\rm h.c.},\nonumber\\
\hat H_{\rm SDW}&{=}\sum_{\alpha=1}^3\sum_{l_1l_2\bm{k} \sigma\sigma'}\!\!
\left[{\bm \Delta}_{\alpha}^{(s)}{\cdot}{\bm \sigma}_{\sigma\sigma'}\right]
\hat c^{\dagger}_{l_1\bm{k}\sigma}
\xi^{(s)}_{l_1l_2}(\bm{Q}_\alpha)
\hat c_{l_2\bm{k}-\bm{Q}_\alpha \sigma'}+{\rm h.c.}.
\end{align}
Here ${\bm \sigma}$ is the vectorial Pauli matrix $\left(\sigma^{(x)},\ \sigma^{(y)},\ \sigma^{(z)}\right)$, and $\bm \Delta_{\alpha}^{(s)}$ $\left(\Delta_\alpha^{(c)}\right)$ is the global amplitude of the $\alpha$-th vectorial SDW (scalar CDW) order parameter determined by the interaction strength via the following MF energy minimization.

Firstly, let's write down the total MF- Hamiltonians describing the two ordered phases
\begin{subequations}\label{HMFSC}
\begin{align}
\hat H_{\rm MF-CDW}&=\hat H_{\rm TB}+\hat H_{\rm CDW},\\
\hat H_{\rm MF-SDW}&=\hat H_{\rm TB}+\hat H_{\rm SDW}.
\end{align}
\end{subequations}
After diagonalizing the two Hamiltonians, we obtain their ground states $\left|\text{CDW-MF}\right\rangle$ and $\left|\text{SDW-MF}\right\rangle$. Secondly, the two MF energies are represented by the expectation values of the original Hamiltonian (\ref{H_Hubbard}) in the two ground states, i.e.,
\begin{subequations}\label{E_VDW_SDW}
\begin{align}
E_{\text{CDW-MF}}&=\left\langle \text{CDW-MF}\left|H\right|\text{CDW-MF}\right\rangle,\\
E_{\text{SDW-MF}}&=\left\langle \text{SDW-MF}\left|H\right|\text{SDW-MF}\right\rangle.
\end{align}
\end{subequations}
Note that the Wick's decomposition procedure is adopted in calculating the above two expectation values. Finally, tuning the SDW or CDW order parameters $\bm \Delta_{\alpha}^{(s)}$ or $\Delta_\alpha^{(c)}$ so that the above two MF- energies are minimized, after which we obtain these order parameters.

An important property of the DW orders obtained at $U$ slightly larger than $U_c$ is that they are either intra-valley orders or inter-valley ones, but not their mixing. To clarify this point, we put aside the sublattice and spin indices of $\chi^{(s)}$ or $\chi^{(c)}$ defined in Eq.~\eqref{SUS} and only focus on the valley degree of freedom, which leads to
\begin{align}
\chi^{(s,c)v_1v_2}_{v_3v_4}\equiv \left \langle T_{\tau }\hat c^{\dagger}_{v_1}(\tau)\hat c_{v_2}(\tau)\hat c^{\dagger}_{v_4}(0)\hat c_{v_3}(0) \right \rangle,
\end{align}
with the valley index $v_i =\pm$ denoting $K$ and $K'$ valleys, respectively. Since the valley-U(1) symmetry of the system requires the conservation of the total value of valleys, i.e. $v_1+v_4=v_2+v_3$, $\chi^{(s,c)v_1v_2}_{v_3v_4}$ should take the form of
\begin{align}\label{block}
\chi^{(s,c)v_1v_2}_{v_3v_4} = \begin{pmatrix}
\chi^{++}_{++} &0  & 0 &\chi^{++}_{--} \\
0 & \chi^{+-}_{+-} & 0  & 0\\
0 & 0 & \chi^{-+}_{-+} & 0 \\
 \chi^{--}_{++}& 0 &0  & \chi^{--}_{--}
\end{pmatrix}.
\end{align}
Here the correspondence between the value of $v_1v_2$ or $v_3v_4$ and the row or column index is $++:1, +-:2, -+:3, --:4$. Due to the block-diagonalized character of the matrices $\chi^{(s,c)}$ shown in Eq.~\eqref{block},  any of their eigenvectors $\xi$ can either take the form of $(a,\ 0,\ 0,\ b)^T$ or of $(0,\ c,\ d,\ 0)^T$. While the form represents the intra-valley order, the latter denotes the inter-valley one, which do not mix. Note that the FS-nesting vectors $\bm{Q}_{\alpha}$ shown in Fig.~\ref{band}(c) and (d) always connect the FSs from different valleys, we can easily conjecture that the induced DW orders are inter-valley orders, which is consistent with our following calculation results.

Note that although the DW order obtained in the charge channel breaks the translational symmetry, the distribution of the charge density in this state is actually translational invariant due to its inter-valley coherence character. Therefore, \sout{it's}{\color{red}it is} inappropriate to name this state as CDW. Instead, it should better be dubbed as the valley DW (VDW), as it breaks the valley-U(1) symmetry. In the following, we rename such quantity as $U_c^{(c)}$, $\Delta_\alpha^{(c)}$ and $\xi^{(c)}(\bm{Q})$ to be $U_c^{(v)}$, $\Delta_\alpha^{(v)}$ and $\xi^{(v)}(\bm{Q})$. The DW order obtained in the spin channel breaks the translational symmetry, the spin-SU(2) and valley-U(1) symmetry. Therefore, we should better name it as valley-spin DW. In the following, we simply dub it as SDW for convenience.

When both $U<U_c^{(s)}$ and $U<U_c^{(v)}$ are satisfied, an effective pairing interaction vertex $V^{\alpha\beta}(\bm{k},\bm{k}')$ is developed through exchanging the short-ranged spin (charge) fluctuations between a Cooper pair. The detailed expression of $V^{\alpha\beta}(\bm{k},\bm{k}')$ is provided in the Appendix \ref{RPAapproach}. It leads to the following linearized gap equation near the superconducting critical temperature $T_c$,
\begin{align}\label{gapeq}
-\frac{1}{(2\pi)^2}\sum_{\beta}\oint_{FS}
dk'_{\Vert}\frac{V^{\alpha\beta}(\bm{k},\bm{k}')}
{v^{\beta}_{F}(\bm{k}')}\Delta_{\beta}(\bm{k}')=\lambda
\Delta_{\alpha}(\bm{k}),
\end{align}
where $\alpha$ and $\beta$ label the bands that cross the FS, corresponding to combined $(mv)$ in Eq.~\eqref{HTB}. $v^{\beta}_F(\bm{k}')$ gives the Fermi velocity and $k'_{\parallel}$ is the tangent component of $\bm{k}'$ along the FS. After discretization, the equation \eqref{gapeq} presents as an eigenvalue problem. The eigenvector $\Delta_{\alpha}(\bm{k})$ represents the gap form factor and the eigenvalue $\lambda$ determines the $T_c$ through $T_c\propto e^{-1/\lambda}$. Symmetry analysis requires that each $\Delta_{\alpha}(\bm{k})$ is attributed to one of the three irreducible representations of the point group $D_3$. Further considering the parity of $\Delta_{\alpha}(\bm{k})$ in the absence of spin-orbit-coupling, there are six possible pairing symmetries \cite{Fanyang}, i.e., $s$, $\left(d_{x^2-y^2},\ d_{xy}\right)$, and $f_{x(x^2-3y^2)}{*}f^{\prime}_{y(y^2-3x^2)}$ pairings for the spin singlet and $\left(p_{x},\ p_{y}\right)$, $f_{x(x^2-3y^2)}$, and $f^{\prime}_{y(y^2-3x^2)}$ pairings for the spin triplet.

Since the superconducting critical temperature $T_c$ is much lower than the total band width of the low-energy emergent flat bands, it is reasonable to only consider the weak-pairing limit, in which only the electrons on the FS participate in the pairing. In such a condition, the Anderson's theorem requires that the Cooper pairing can only take place between inter-valley. Moreover, these inter-valley pairings are neither valley-singlet pairing nor valley-triplet one, but instead are a mixing between them, as the square of the total vectorial valley of the Cooper pair is not a good quantum number here. Actually, if an electron with momentum-valley $\bm{k}$-$K$ is on the FS and thus can participate in the pairing, the electron with momentum-valley $\bm{k}$-$K'$ is generally away from the FS and thus cannot participate in the pairing, which leads to a ratio of 1:0 between the amplitudes for the parings of $c^{\dagger}_{\bm{k}K} c^{\dagger}_{\bm{-k}K'}$ and $c^{\dagger}_{\bm{k}K'} c^{\dagger}_{\bm{-k}K}$, leading to a 1:1 mixing between the valley-singlet and valley-triplet pairings.

\section{Chiral SO(4)-DW and degenerate SC at $J_H=0$}
\label{SecWithoutJH}
As introduced in Sec.~\ref{secModel}, when the inter-valley Hund's coupling is neglected, the MA-TBG has an SU(2)$_K\times$SU(2)$_{K^\prime}$ symmetry, with each valley independently hosting a spin-SU(2) symmetry. In this section, we will explore the consequence of such a symmetry. It will be seen below that degeneracies will take place either between the SDW and VDW or between the singlet and triplet SCs. The degeneracy between the SDW and VDW orders, in combination with the three-folded degeneracy among the wave vectors of the DW orders caused by the $D_3$ point group of the MA-TBG, would make them mix into a chiral SO(4) DW order. A series of intriguing properties of this chiral SO(4) DW state are studied.

\begin{figure}
\centering
\includegraphics[width=0.48\textwidth]{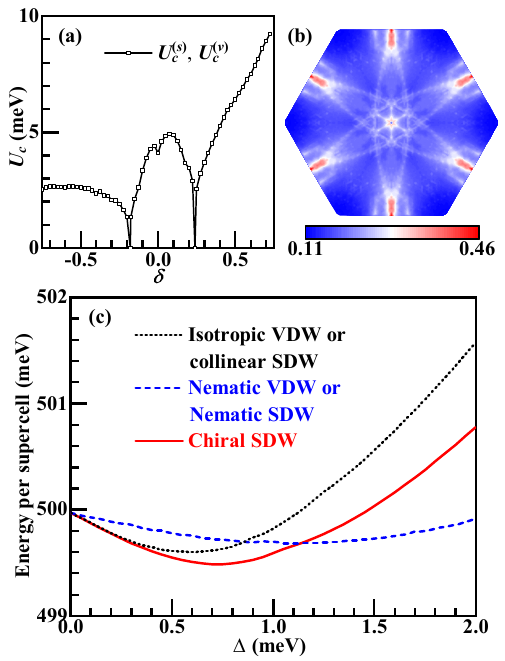}
\caption{(a) Doping dependence of $U^{(s)}_{c}$ and $U^{(v)}_{c}$. (b) Distribution of $\chi(\bm{q})$ in the MBZ for $\delta=0.240$, corresponding to the e-VHS in Fig. \ref{band}.  (c)  The energies of MF states determined by $H_{\rm MF-SDW}$ and $H_{\rm MF-VDW}$ for several different configurations at the e-VHS point with $U=4$ meV . The non-zero order parameters are $\Delta_1^{(v)}= \Delta_2^{(v)}= \Delta_3^{(v)} = \Delta$ for the isotropic VDW, $\Delta^{(s)}_{1,z}=\Delta^{(s)}_{2,z}=\Delta^{(s)}_{3,z} = \Delta $ for the collinear SDW, $\Delta_{1}^c = \Delta$ for the nematic VDW, $\Delta^{(s)}_{1,z} = \Delta$ for the nematic SDW, and $\Delta^{(s)}_{1,x}=\Delta^{(s)}_{2,y}=\Delta^{(s)}_{3,z} = \Delta$ for the chiral SDW, in which the energies of the isotropic and nematic VDWs are exactly equal to those of the collinear and nematic SDW, respectively. These five configurations take the minimal energies of 499.603 meV, 499.603 meV, 499.681 meV, 499.681 meV, and 499.484 meV, respectively, when their $\Delta$ take 0.602 meV, 0.602 meV, 1.131 meV, 1.131 meV, and 0.720 meV.}
\label{DW_Orders}
\end{figure}

\subsection{Degenerate DW Orders Mixing into SO(4) DW}
\label{DegeneracyDW}

The doping dependence of the critical interaction strengths $U^{(s)}_{c}$ and $U^{(v)}_{c}$ are shown in Fig. \ref{DW_Orders}(a). From Fig. \ref{DW_Orders}(a), the $U^{(s)}_{c}$ and $U^{(v)}_{c}$ are at the order of the band width of the flat band. Two features are obvious in Fig. \ref{DW_Orders}(a). The first feature is that both $U^{(s)}_{c}$ and $U^{(v)}_{c}$ go to zero at the two VH dopings, suggesting that an infinitesimal interaction would drive DW orders at these dopings. This feature originates from the fact that the divergent DOS together with the good FS nesting makes even the bare susceptibility $\chi^{(0)}$ diverge. The second feature is that the $U^{(s)}_{c}$ and $U^{(v)}_{c}$ are exactly equal for a large doping range around the VH dopings. Further more, the eigenvectors $\xi^{(s)}$ and $\xi^{(v)}$ corresponding to the largest eigenvalues of $\chi^{(s)}(i\omega=0)$ and $\chi^{(c)}(i\omega=0)$ are identical too, which take the form of $(0,\ c,\ d,\ 0)^T$ and belong to the inter-valley type of DW orders, originating from the inter-valley FS-nesting shown in Figs.~\ref{band}(c) and 1(d). Such a degeneracy originates from the SU(2)$_K\times$SU(2)$_{K^\prime}$ symmetry of the MA-TBG system, as clarified below.

Due to the SU(2)$_K\times$SU(2)$_{K^\prime}$ symmetry of MA-TBG in the case of $J_H=0$, we can define the unitary symmetry operation $\hat P: c_{i}\to\hat Pc_{i}\hat P^{\dagger}$ with the following explicit formula,
\begin{align}\label{unitary_symmetry}
\hat c_{i+\uparrow}  \rightarrow \hat c_{i+\uparrow},\ \ \hat c_{i+\downarrow}  \rightarrow  \hat c_{i+\downarrow}, \ \
\hat c_{i-\uparrow}  \rightarrow \hat c_{i-\uparrow},\ \ \hat c_{i-\downarrow}  \rightarrow -\hat c_{i-\downarrow}.
\end{align}
One can easily check $\left[\hat {P},\ \hat {H}\right]=0$ from Eq. \eqref{H_Hubbard} (set $J_H=0$) and Eq. \eqref{unitary_symmetry}. A consequence of this symmetry is that it maps an inter-valley VDW order to the $z$-component of an inter-valley SDW (abbreviated as the z-SDW) one with the same wave vector $\bm{Q}$ and form factor $\xi_{v_1v_2}(\bm Q)$, i.e.,
\begin{subequations}
\begin{align}
\hat{O}_{\rm VDW}&\equiv\sum_{\iota_1v_1,\iota_2v_2,\bm{k}\sigma}\hat c^{\dagger}_{\iota_1v_1 \bm{k} \sigma}\xi_{\iota_1v_1,\iota_2v_2}(\bm Q)\hat c_{\iota_2v_2 \bm{k}-\bm{Q}\sigma},\\
\hat{O}_{\rm z-SDW}&\equiv \!\sum_{\iota_1v_1,\iota_2v_2,\bm{k}\sigma\sigma'} \!\!\!\! \hat c^{\dagger}_{\iota_1v_1 \bm{k}\sigma}\xi_{\iota_1v_1,\iota_2v_2}(\bm Q)\sigma^z_{\sigma\sigma'} \hat c_{\iota_2v_2 \bm{k}-\bm{Q}\sigma'},\label{OzSDW}
\end{align}
\end{subequations}
which satisfy
\begin{align}
\hat P^{\dagger}\hat O_{\rm VDW}\hat P=\hat O_{\rm z-SDW}.
\label{mapDW}
\end{align}
Here the inter-valley condition for the DW orders requires
\begin{align}\label{inter_valley_xi}
\xi_{\iota_1v_1,\iota_2v_2}=\delta_{\bar{v}_1,v_2}\xi_{\iota_1v_1,\iota_2\bar{v}_1}
\end{align}
One can easily check Eq. \eqref{mapDW} by using Eq. \eqref{unitary_symmetry} and Eq. \eqref{inter_valley_xi}.

Now let's gradually enhance the interaction strength $U$ from zero and monitor the formation of the VDW and SDW orders. Initially, $U$ is so small that the formation of neither the SDW nor the VDW can gain energy, and thus no DW orders are formed. On the one hand, supposing at the critical interaction strength $U_c^{(v)}$, the formation of a VDW order with a wave vector $\bm{Q}$ and a form factor $\xi^{(v)}(\bm{Q})$ begins to gain energy. Then from the mapping in Eq.~\eqref{mapDW} and the fact of $\left[\hat {P},\ \hat {H}\right]=0$, \sout{it's}{\color{red}it is} easily proved that the formation of a z-SDW order with the same wave vector and form factor can also gain energy because
\begin{eqnarray}\label{E_VDW_zSDW}
E_{\rm VDW}&=&\left\langle {\rm VDW}\left| \hat H \right|{\rm VDW}\right\rangle=\left\langle {\rm VDW}\left|\hat P^{\dagger} \hat H \hat P\right|{\rm VDW}\right\rangle\nonumber\\&=&\left\langle {\rm z{-}SDW}\left| \hat H \right|{\rm z{-}SDW}\right\rangle=E_{\rm z-SDW}.
\end{eqnarray}
Therefore, we have $U_c^{(v)}\ge U_c^{(s)}$. On the other hand, let's suppose $U$ is enhanced to $U_c^{(s)}$ so that the formation of an SDW order with an arbitrary direction of magnetization with a wave vector $\bm{Q}$ and form factor $\xi^{(s)}(\bm{Q})$ begins to gain energy. Note that from the spin-SU(2) symmetry, we can always rotate the direction of the magnetization to the $z$-axis without costing energy, thus $U_c^{(s)}$ is also the critical $U$ for the z-SDW order. As for arbitrary $U>U_c^{(s)}$, the formation of a z-SDW state can gain energy, then from Eq. \eqref{E_VDW_zSDW} the formation of a VDW state can also gain energy, suggesting $U_c^{(v)}\le U_c^{(s)}$. The combination of both hands leads to $U_c^{(v)}=U_c^{(s)}\equiv U_c$, and the wave vector $\bm{Q}$ together with the form factor $\xi(\bm{Q})$ of both DW orders should be identical.

On the above we prove the degeneracy between the SDW and the VDW. Due to this degeneracy, the two DW order parameters will generally be mixed to lower the energy. In the Appendix.~\ref{SO4} we study how they would be mixed via combined Ginzburg-Landau (G-L) theory and the microscopic calculations. As a result, our results yield that the two DW orders should be mixed with a $\pi/2$ phase difference, suggesting that the MF Hamiltonian involving both orders is
\begin{align}\label{HMF0}
\hat H_{\rm MF-DW}&=\hat H_{\rm TB} +\sum_{\iota_1 \iota_2 \bm{k}\sigma\sigma'}\left(\Delta^{(v)}\delta_{\sigma\sigma'}+i\bm{\Delta}^{(s)}\cdot\bm{\sigma}_{\sigma\sigma'}\right)\nonumber\\
&\qquad \times c^{\dagger}_{\iota_1 \bf{K}\bm{k}\sigma}\xi_{\iota_1 \bf{K} \iota_2 \bf{K'}}(\bm{Q})
\hat c_{\iota_2 \bf{K'}\bm{k}-\bm{Q} \sigma'}+{\rm h.c.}
\end{align}
In this form of DW ordered state, the SU(2)$_K\times$SU(2)$_{K^\prime}$ symmetry of the system would be embodied as the SO(4) symmetry for the DW order parameters.

\subsection{Consequence of degeneracy among wave vectors}
\label{ConsequenceDegeneracy}

On the above, we have proved the degeneracy between the SDW and VDW orders at the critical point. Note that only one single wave vector $\bm{Q}$ of the DW orders is considered. In such a case, the degeneracy not only applies at the critical point but also at any $U>U_c$: the ground-state energies of both DW states are always equal to each other due to Eq. \eqref{E_VDW_zSDW} and the spin-SU(2) symmetry. However, for the MA-TBG, there is a three-folded rotational symmetry, which brings about three degenerate wave vectors for the DW orders simultaneously. In such a case, the DW components hosting these degenerate wave vectors can be mixed, leading to a different situation: the degeneracy between SDW and VDW only applies at $U=U_c$, but not at $U>U_c$ where the ground-state energy of the SDW state with mixed wave vectors is lower than that of the VDW state, as will be discussed below.

As shown in Figs.~\ref{band}(c) and \ref{band}(d), the FS of MA-TBG exhibits three-folded degenerate nesting vectors $\bm{Q}_{\alpha} (\alpha=1,\ 2,\ 3)$, which in the weak-coupling treatment are just the three degenerate wave vectors of the DW orders. This point is supported by the distribution of the largest eigenvalue $\chi(\bm{q})$ of the bare susceptibility matrix at $i\omega=0$ in the MBZ, as shown in Fig.~ \ref{DW_Orders} (b) is for the e-VH doping. Figure~\ref{DW_Orders}(b) exhibits a six-folded symmetric pattern peaking at $\pm\bm{Q}_\alpha(\alpha=1,2,3)$. As the three $\bm{Q}_\alpha$ are near the three $M_{\alpha}$-points in the MBZ, we just set $\bm{Q}_\alpha=M_{\alpha}$ for simplicity. When interactions turn on, the spin or charge susceptibilities first diverge at the three $\bm{Q}_\alpha$, yielding the three degenerate wave vectors as $\bm{Q}_\alpha$.

In the presence of degenerate wave vectors, the degeneracy between SDW and VDW orders is still tenable at the critical point, including the relations $U_c^{(v)}=U_c^{(s)}$ and $\xi^{(v)}(\bm{Q}_\alpha)=\xi^{(s)}(\bm{Q}_\alpha)$. The reason for this degeneracy is clear in the framework of RPA: the critical interaction $U_c^{(s)}$ or $U_c^{(v)}$ is determined by the condition that the denominator matrix in Eq.~\eqref{Renorm_SUS_spin} or Eq.~\eqref{Renorm_SUS_charge} begins to have zero eigenvalue at some $\bm{q}$. In the presence of degenerate wave vectors, this condition is first satisfied by the three degenerate momenta simultaneously, which means that the condition $U=U_c^{(v,s)}$ is also the condition that the formation of the VDW or SDW orders with any one of the three wave vectors can first gain energy. Therefore the above energy-based proof for the single-$\bm{Q}$ case also applies here.

However, the degeneracy between the SDW and VDW orders is broken for a general $U>U_c^{(v)}=U_c^{(s)}$, wherein the interaction among the degenerate order-parameter components corresponding to the degenerate wave vectors energetically favors the SDW. The mixing of the three degenerate components of the VDW and SDW orders leads to the order-parameter fields given by Eq.~\eqref{HDW}. From the formula of $\hat{P}$ defined in Eq. \eqref{unitary_symmetry}, \sout{it's}{\color{red}it is} easily checked that a VDW state formed by the mixing of three degenerate components with wave vectors $\bm{Q}_\alpha$, form factors $\xi(\bm{Q}_\alpha)$, and global amplitude $\Delta_{\alpha}$, is described by
\begin{align}\label{HVDW0}
\hat H_{\rm VDW}&=\sum_{\alpha=1}^3\!\sum_{l_1l_2\bm{k} \sigma}\Delta_\alpha
\hat c^{\dagger}_{l_1\bm{k}\sigma}
\xi_{l_1l_2}(\bm{Q}_\alpha)
\hat c_{l_2\bm{k}-\bm{Q}_\alpha \sigma}{+}{\rm h.c.},
\end{align}
we have
\begin{align}
\hat P^{\dagger}\hat H_{\rm VDW}\hat P=\hat H_{\rm col-SDW}.
\label{mapDWH}
\end{align}
with
\begin{align}
\hat{H}_{\rm col-SDW}&\equiv \sum_{\alpha=1}^3\!\sum_{l_1l_2\bm{k} \sigma_1\sigma_2}\Delta_\alpha \sigma^z_{\sigma_1\sigma_2}
\hat c^{\dagger}_{l_1\bm{k}\sigma_1}
\xi_{l_1l_2}(\bm{Q}_\alpha)
\hat c_{l_2\bm{k}-\bm{Q}_\alpha \sigma_2}{+}{\rm h.c.}.\label{OcolSDW}
\end{align}
Obviously, the $\hat{H}_{\rm col-SDW}$ defined above is a special case of the $\hat{H}_{\rm SDW}$ defined in Eq. \eqref{HDW} with setting $\xi^{(s)}=\xi$ and $\bm{\Delta}_{\alpha}=\Delta_{\alpha}\bm{e}_z$. In such an SDW state, all the three degenerate vectorial SDW components are along the same $z$-direction, forming a collinear SDW state. Therefore, in the presence of degenerate wave vectors, the SU(2)$_K\times$SU(2)$_{K^\prime}$ symmetry of the MA-TBG maps any inter-valley VDW order into an inter-valley collinear SDW order with the same wave vector and form factor, and hence both DW states share the same ground-state energy. However, the general form of SDW states given in Eq.~\eqref{HDW} not only includes the collinear SDW states but also includes the non-collinear ones. Therefore, the ground-state energy of the SDW state is at least no higher than that of the VDW state in the presence of degenerate wave vectors. Our numerical calculations shown below single out the non-coplanar chiral SDW state to be the SDW state with the lowest energy, which, of course, is lower than that of the VDW state.

To find the energetically most favored DW state, we should take the three (nine) components of the VDW (SDW) order parameter, $\Delta_{\alpha}^{(v)}\ (\alpha=1,2,3)$ $\left(\Delta_{\alpha,\mu}^{(s)}\ (\alpha=1,2,3;\ \mu=x,y,z)\right)$ in Eq. \eqref{HDW} as the variational parameters to minimize the energy of the Hamiltonian \eqref{H_Hubbard} in the VDW (SDW) MF state generated by the MF Hamiltonian \eqref{HMFSC}.

Before performing the energy minimization, \sout{it's}{\color{red}it is} helpful to classify all the possible configurations of the VDW and the SDW order parameters from the G-L theory. The G-L theory provided in the Appendix.~\ref{puresdwvdw} suggests that there exist three SDW configurations, i.e. the collinear SDW state, the chiral SDW state and the nematic SDW state. In the collinear state, the three SDW order parameters $\bm{\Delta}_{1}=\bm{\Delta}_{2}=\bm{\Delta}_{3}$. In the chiral state, they satisfy $\bm{\Delta}_{1}\perp  \bm{\Delta}_{2} \perp  \bm{\Delta}_{3}$ and $|\bm{\Delta}_{1}|=  |\bm{\Delta}_{2}|=|\bm{\Delta}_{3}|$. In the nematic state, only one of the three $\bm{\Delta}_{\alpha}(\alpha=1, 2, 3)$ exists, and the other two vanishes. As for the VDW, there exist two possible configurations, i.e. the isotropic-VDW state and the nematic-VDW state. While the former contains three VDW components with equal amplitude for the three wave vectors, the latter only contains one for one arbitrarily chosen wave vector.

For the VDW states, our numerical results yield that the energetically most favored state is the isotropic VDW state with $\Delta_{1}^{(v)}=\Delta_{2}^{(v)}=\Delta_{3}^{(v)}=\Delta$. The energy of this state is exactly equal to that of the collinear-SDW state with $\Delta_{\alpha,z}^{(s)}=\Delta; \Delta_{\alpha,x/y}^{(s)}=0$, as proved on the above. To compare, we also calculate the energy of the nematic VDW state with only $\Delta_{1}^{(v)}=\Delta$ as the nonzero component, whose energy is exactly equal to the nematic SDW state with only $\Delta_{1,z}^{(s)}=\Delta$ as the nonzero component. The $\Delta$-dependence of the two VDW states (and the associate SDW states) are shown in Fig.~\ref{DW_Orders}(c), which verifies the isotropic VDW state as the energetically most favored VDW state, consistent with the so called 3Q VDW state defined in Ref.~\cite{LiangFu}. However, this 3Q-VDW state is beaten by the non-coplanar chiral SDW state with $\Delta_{1,x}^{(s)}=\Delta_{2,y}^{(s)}=\Delta_{3,z}^{(s)}=\Delta$ as the nonzero components, which is among the energetically most favored degenerate SDW states, consistent with Ref. \cite{Fanyang}. These degenerate ground states are related by the spin-SU(2) rotations. In each of these degenerate lowest-energy SDW states, the three SDW order-parameter components $\bm\Delta^{(s)}_\alpha$ with equal amplitudes satisfy $\bm\Delta^{(s)}_1\perp\bm\Delta^{(s)}_2\perp\bm\Delta^{(s)}_3$, leading to non-coplanar structure with spin chirality. Such chiral SDW states cannot be mapped to any VDW state by the SU(2)$_K\times$SU(2)$_{K^\prime}$ symmetry operation. The $\Delta$-dependence of the energy of the chiral SDW states is compared to that of the VDW states in Fig.~\ref{DW_Orders} (c), which verifies that the former is energetically more favored than the latter.

\subsection{Chiral SO(4) Spin-Valley DW}
\label{SCDegeneracy}

As clarified in the above two subsections, although the SU(2)$_K\times$SU(2)$_{K^\prime}$ symmetry brings about the degeneracy between the SDW and VDW orders at the critical point $U=U_c$, the SDW order with a non-coplanar chiral spin configuration wins over the VDW at the ground state for general realistic $U>U_c$. However, the SU(2)$_K\times$SU(2)$_{K^\prime}$ symmetry still plays an important role in determining the ground state in general cases. Assuming that the chiral SDW state with $\Delta_{1,x}^{(s)}=\Delta_{2,y}^{(s)}=\Delta_{3,z}^{(s)}=\Delta$ obtained above is the ground state, let's perform the symmetry operation $\hat{P}$ on this state. Consequently, we obtain a DW state with two vectorial SDW components pointing toward the $x$- and $y$-directions mixed with one scalar VDW component. This state would have the same energy as the chiral SDW state. This fact tells us that the ground state of the system is generally a mixing between the SDW and VDW orders. As clarified in Sec.\ref{DegeneracyDW}, in the case of one single wave vector, the SDW and VDW would be mixed in the manner of a $\pi/2$ phase difference to form the SO(4) DW. When all the three SO(4) DW components for the three wave vectors turn on, the general form of the MF Hamiltonian for the DW state reads,
\begin{align}\label{HMF}
\hat H_{\rm MF-DW}&=\hat H_{\rm TB} +\sum_{\alpha=1}^3\sum_{l_1l_2\bm{k}\sigma\sigma'}\left(\Delta_{\alpha}^{(v)}\delta_{\sigma\sigma'}+i\bm{\Delta}_{\alpha}^{(s)}\cdot\bm{\sigma}_{\sigma\sigma'}\right)\nonumber\\
&\qquad \times c^{\dagger}_{l_1\bm{k}\sigma}\xi_{l_1l_2}(\bm{Q}_\alpha)
\hat c_{l_2\bm{k}-\bm{Q}_\alpha \sigma'}+{\rm h.c.}\nonumber\\
& = \hat H_{\rm TB} + \sum_{\alpha=1}^3\sum_{l_1l_2\bm{k}\sigma\sigma'}\left[\left(\bm{\varDelta}_\alpha\cdot \bm\varSigma_\alpha\right)_{\sigma\sigma'}\right. \nonumber\\
&\qquad \times \left.\hat c^{\dagger}_{l_1\bm{k}\sigma}\xi_{l_1l_2}(\bm{Q}_\alpha)
\hat c_{l_2\bm{k}-\bm{Q}_\alpha \sigma'}+{\rm h.c.}\right],
\end{align}
where the 4-component vector $\bm{\varDelta}_{\alpha}\equiv\left(\Delta^{(v)}_{\alpha}, \bm\Delta^{(s)}_{\alpha}\right)=\left(\Delta^{(v)}_{\alpha}, \Delta^{(s)}_{\alpha,x},\Delta^{(s)}_{\alpha,y},\Delta^{(s)}_{\alpha,z}\right)\in\mathbb{R}^4$ and $\bm\varSigma_{\alpha}=\left(\sigma^{(0)},i\bm\sigma\right)$ with $\sigma^{(0)}$ to be the $2\times2$ identity matrix. Here we have totally twelve variational parameters $\bm{\varDelta}_{\alpha}(\alpha=1, 2, 3)$.

\begin{figure*}[htbp]
\centering
\includegraphics[width=0.9\textwidth]{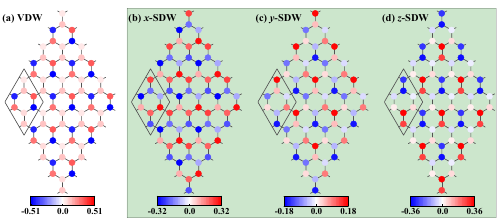}
\caption{The real-space distributions of the scalar inter-valley charge density (a) and the three components of the vectorial inter-valley spin density (b-d) for a typical ground state configuration with $\bm \varDelta_1=(0.47, -0.19, -0.22, 0.46)$ meV, $\bm \varDelta_2=(-0.49, 0.13, -0.11, 0.50)$ meV, and $\bm \varDelta_3=(-0.24, -0.64, -0.19, -0.11)$ meV in the chiral SO(4) DW phase for $J_H=0$.}\label{SO4DW}
\end{figure*}

Before performing the energy minimization, we have done a G-L theory based analysis in the Appendix.~\ref{mixing_3Q} to classify the possible configurations of the three SO(4) DW order parameters as possible solutions to minimize the G-L free-energy function. Consequently, only three possible solutions exist, i.e. the collinear SO(4) spin-valley DW state, the chiral SO(4) spin-valley DW state and the nematic SO(4) spin-valley DW state. In the collinear state, the three DW order parameters $\bm{\varDelta}_{1}=
\bm{\varDelta}_{2}=\bm{\varDelta}_{3}$. In the chiral state, they satisfy $\bm{\varDelta}_{1}\perp  \bm{\varDelta}_{2} \perp  \bm{\varDelta}_{3}$ and $|\bm{\varDelta}_{1}|=  |\bm{\varDelta}_{2}|=|\bm{\varDelta}_{3}|$. In the nematic state, only one of the three $\bm{\varDelta}_{\alpha}(\alpha=1, 2, 3)$ exists, and the other two vanish.

Our energy-minimization result yields that the chiral SO(4) spin-valley DW states are the ground states of the system. These states include the chiral SDW with $\bm{\varDelta}_{1}=(0,\Delta,0,0),\bm{\varDelta}_{2}=(0,0,\Delta,0),\bm{\varDelta}_{3}=(0,0,0,\Delta)$ as a special example. However, there are simultaneously many other degenerate ground states with equal energy to this state, forming a ground-states set. This set of states are obtained through performing all the possible global SO(4)-rotations on the three $\bm{\varDelta}_{\alpha}$ of the chiral SDW state within the $\mathbb{R}^4$ parameter space. Such a ground-state degeneracy results from the spontaneous breaking of the SO(4) symmetry which originates from the physical SU(2)$_K\times$SU(2)$_{K^\prime}$ symmetry, see Appendix \ref{SO4}. Therefore, the ground state of the MA-TBG should be a mixing between the SDW and VDW with a particular manner: this DW state possesses three coexisting wave vectors $\bm{Q}_\alpha$, with each $\bm{Q}_\alpha$ distributed to a 4-component DW order parameter which comprises of one VDW component and three SDW ones. The three 4-component vectorial DW order parameters with equal amplitude are perpendicular to each other and can globally arbitrarily rotate in the $\mathbb{R}^4$ parameter space. We call such a DW state as the Chiral SO(4) spin-valley DW. Besides, as the obtained inter-valley DW states break the valley-U(1) symmetry, the valley-U(1) rotation about the valley $\tau_z$-axis will rotate the DW order parameters in the valley $(\tau_x,\tau_y)$ plane. Concretely, it will change the form factor $\xi$ in Eq. (\ref{HMF}) by a multiplied phase factor $e^{i\alpha}$. Such valley-U(1) rotation brings about extra ground-state degeneracy.

The Goldstone-modes fluctuations grown on top of the chiral SO(4) DW ground state are intriguing, considering the continuous SO(4) and valley-U(1) symmetry-breaking, combined with the wave-vector degeneracy. Firstly, let's globally rotate the three $\bm{\varDelta}_\alpha$ so that one of it, say $\bm{\varDelta}_1$ is rotated from its polarization direction to the three remaining perpendicular directions in the $\mathbb{R}^4$ space, and $\bm{\varDelta}_{2, 3}$ are also operated by these global rotations. Such global rotations lead to three gapless Goldstone modes. Secondly, let's choose the global rotation manner so that $\bm{\varDelta}_1$ is fixed unchanged, and $\bm{\varDelta}_2$ can freely rotate toward the two remaining directions under the condition $\bm{\varDelta}_1\perp\bm{\varDelta}_2$, leading to two more gapless Goldstone modes. Thirdly, let's fix the rotation plane to be that expanded by $\bm{\varDelta}_1$ and $\bm{\varDelta}_2$, under which the $\bm{\varDelta}_3$ can only rotate toward the remaining one direction under the condition $\bm{\varDelta}_3\perp\bm{\varDelta}_1\perp\bm{\varDelta}_2$, leading to one more Goldstone mode. Finally, the continuous valley-U(1) symmetry breaking brings about another gapless Goldstone mode, which is the rotation of the order parameters in the valley $(\tau_x,\tau_y)$ plane. All together, we have seven branches of gapless Goldstone modes, much more than those in conventional SDW states. For example, the Neel SDW state on the square or honeycomb lattice has only two branches of gapless acoustic Goldstone modes.

Due to the Mermin-Wagner theorem, at finite temperature, the Goldstone-modes fluctuations in the 2D MA-TBG system would destroy the long-range chiral SO(4) DW order which breaks the continuous SO(4) and valley-U(1) symmetry. However, the short-range fluctuations of this DW order still exist. Further more, there exists a characteristic temperature $T_M$ below which the correlation length of the DW order begins to enhance promptly, and the local environment around an electron is similar with that in the presence of long-range order. As a result, many properties  exhibited in the experiment are also similar with the latter case. It was argued in Ref.~\cite{Xu2} that the SDW-correlated state can explain such experimental results as the transport property at finite temperature. The chiral SO(4) DW state can be obtained from the chiral SDW state through an SU(2)$_K\times$SU(2)$_{K^\prime}$ rotation, which is a unitary transformation and doesn't alter the band structure. Therefore, this SO(4) DW state is also ready to explain similar experimental results. Note that in addition to the continuous SO(4) and valley-U(1) symmetry, the discrete TRS is also broken here, which can possibly maintain at finite temperature, leading into such experimental consequence as the Kerr effect.

The topological properties of the chiral SO(4) DW state might probably be nontrivial with nonzero Chern number. As this state is related to the chiral SDW state through a unitary transformation, the two states share the same topological properties. The chiral SDW states with three degenerate wave vectors have been studied previously in other circumstances\cite{TaoLi,Martin,Kato,Ying}, which suggests that when an SDW gap opens at the Fermi level, this state has a nontrivial topological Chern number and is thus an interaction-driven spontaneous quantum anomalous Hall (QAH) insulator \cite{xidai121, xidai122, ZhangTwist}. Therefore, the chiral SO(4) DW state obtained here might also be a spontaneous QAH insulator, as long as the single-particle gap caused by the DW order opens at the Fermi level. Experimentally, the half-filled MA-TBG is indeed a correlated insulator \cite{caoyuan2}, which thus might probably be a QAH insulator.

\begin{figure}
\centering
\includegraphics[width=0.48\textwidth]{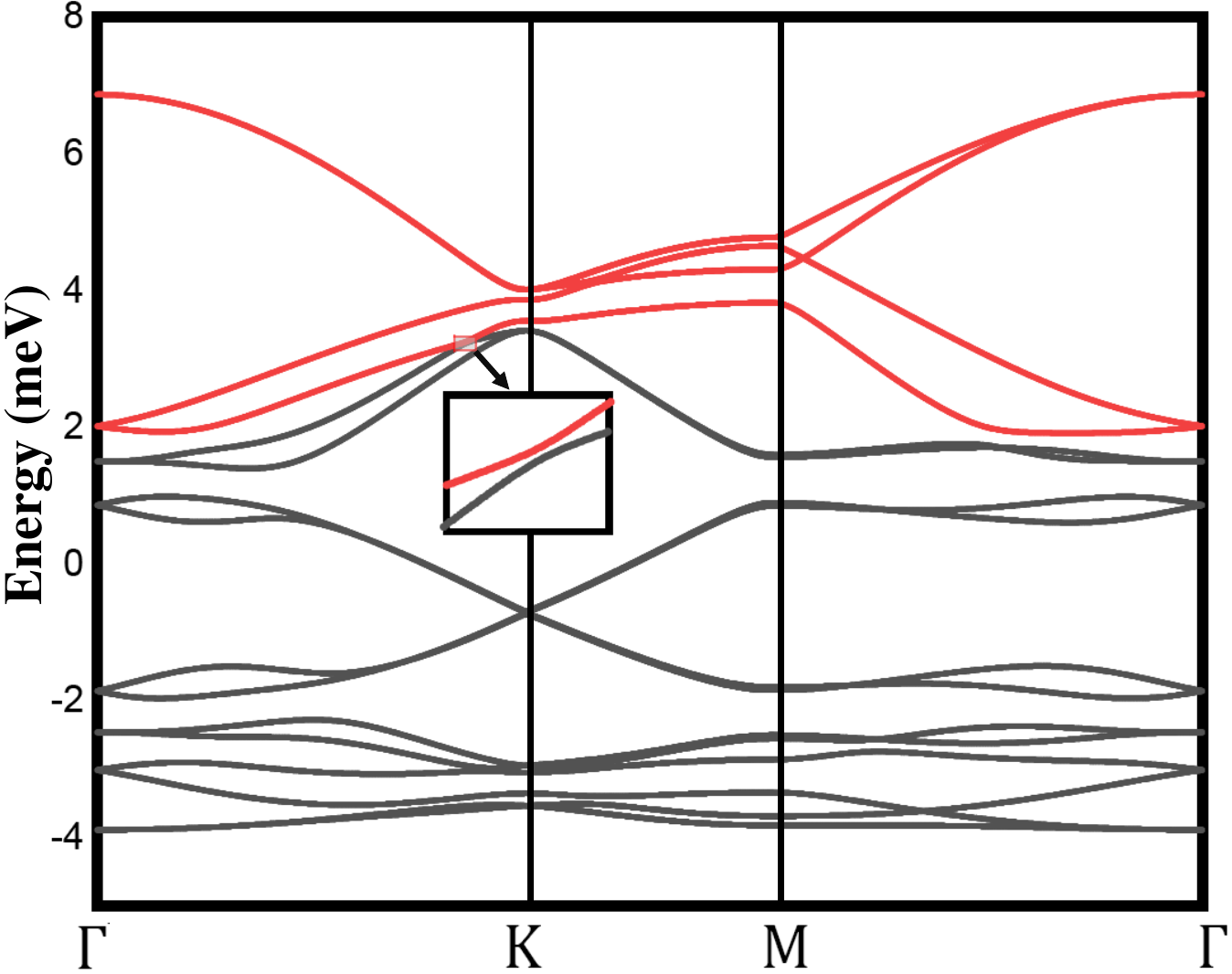}
\caption{The band structure along the high-symmetric lines for the chiral SO(4) DW state at
 half filling in the electron-doped case. The red solid lines and black solid lines represent for the band structures of the conduction bands and valence bands respectively. Inset: the nearly crossing and tiny splitting between the lowest conduction band and the highest valence band.}\label{DW_band}
\end{figure}

In our model, the band structure reconstructed in the chiral SO(4) DW state for the half filling in the electron-doped case is shown in Fig.~\ref{DW_band}. Globally, the conduction bands (red solid) overlap with the valence bands (black solid), leading to a metallic state instead of an insulator. However, as there is no degenerate point in momentum space between the highest valence band and the lowest conduction band, the two bands are separate by a direct gap. In such a case, the total Chern number of the valence bands is still well-defined. The situation for the hole-doped case is similar. Our calculation of the Chern number of the valence bands through the formula provided in Ref.\cite{TaoLi,Martin} yields the number of 4 (-4) for the half filling in the electron-doped (hole-doped) case, suggesting the possibility of QAH effect. Although the DW gap under the present interaction parameters is not large enough to fully separate the valence bands and the conduction bands, they can be fully separated for enhanced interaction parameters, leading to real QAH effect. We leave this topic for future study.

To show the real-space pattern of the chiral SO(4) DW orders, we introduce the following inter-valley site-dependent valley and spin densities defined as
\begin{subequations}\label{VDW_SDW_ORDER}
\begin{align}
\Delta^{(v)}_j&=\left \langle  \hat{c}^{\dagger}_{j+\uparrow }\hat{c}_{j-\uparrow } +\hat{c}^{\dagger}_{j+\downarrow }\hat{c}_{j-\downarrow } + {\rm h.c.}\right \rangle,\\
\Delta^{(s)}_{j,x}&=\left \langle  \hat{c}^{\dagger}_{j+\uparrow }\hat{c}_{j-\downarrow} +\hat{c}^{\dagger}_{j+\downarrow }\hat{c}_{j-\uparrow} + {\rm h.c.} \right \rangle,\\
\Delta^{(s)}_{j,y}&=\left \langle -i \hat{c}^{\dagger}_{j+\uparrow }\hat{c}_{j-\downarrow } +i \hat{c}^{\dagger}_{j+\downarrow }\hat{c}_{j-\uparrow } + {\rm h.c.} \right \rangle,\\
\Delta^{(s)}_{j,z}&=\left \langle  \hat{c}^{\dagger}_{j+\uparrow}\hat{c}_{j-\uparrow } -\hat{c}^{\dagger}_{j-\downarrow }\hat{c}_{j+\downarrow } + {\rm h.c.}\right \rangle.
\end{align}
\end{subequations}
The real-space distributions of these densities are shown in Fig.~\ref{SO4DW} for an arbitrarily chosen ground state with $\bm \varDelta_1=(0.47, -0.19, -0.22, 0.46)$, $\bm \varDelta_2=(-0.49, 0.13, -0.11, 0.50)$ and $\bm \varDelta_3=(-0.24, -0.64, -0.19, -0.11)$. This pattern leads to a $2\times2$-enlarged unit cell as enclosed by the black diamonds in Fig.~\ref{SO4DW}, which contains 8 sites or 16 orbitals. Such a translation-symmetry breaking has not been detected by experiments yet, which might possibly be caused by that the inter-valley valley or spin order in this system can not be easily coupled to conventional experimental observables. Obviously, both the VDW and SDW orders are nematic in the shown configuration, spontaneously breaking the $C_3$ rotational symmetry of the MA-TBG \cite{Chichinadze2019}. However, this state can also arbitrarily rotate to other isotropic states such as the chiral SDW state. Concretely, the orientations of the three $\bm{\varDelta}_\alpha$ can be pinned down by an added infinitesimal term breaking the SU(2)$_K\times$SU(2)$_{K^\prime}$ symmetry, such as an imposed weak magnetic field studied below or a tiny inter-valley Hund's-rule coupling that will be studied in the next section.

To investigate how an imposed infinitesimal magnetic field will pin down the direction of the polarization of the chiral SO(4) DW obtained here through the Zeeman coupling, the following Zeeman term is added into the Hamiltonian \eqref{H_Hubbard},
\begin{eqnarray}
H_{\rm Zeeman}=J_Z\sum_{i,v}\left(\hat c^{\dagger }_{iv\uparrow}\hat c_{iv \uparrow }-\hat c^{\dagger }_{iv\downarrow }\hat c_{iv\downarrow }\right),
\end{eqnarray}
where $J_Z=0.01$ meV is adopted. The energy of $\hat H_{\rm TB}+\hat H_{\rm int}+\hat H_{\rm Zeeman}$ is optimized in the state determined by $H_{\rm MF-DW}$ in Eq.~\eqref{HMF}. Our numerical results for the optimized order parameters are as follow. Firstly, the three relative phase angles between the VDW and SDW orders are $\theta_\alpha\approx\frac{\pi}{2}$, approximately maintaining the SO(4) symmetry. Secondly, among the three DW order parameters $\bm\varDelta_\alpha$, an arbitrarily chosen one, say $\bm\varDelta_1$, takes the form of $\bm\varDelta_1\approx(\Delta, 0,0,0)$, denoting a VDW order, and the remaining two both take the form of $(0,\Delta_1,\Delta_2,0)$ and are perpendicular to each other, denoting two mutually-perpendicular SDW orders polarized within the $xy$-plane. Therefore, we obtain a spin-valley DW ordered state which hosts one scalar VDW order mixed with two mutually perpendicular vectorial SDW orders oriented within the $xy$-plane, with the three DW order parameters randomly distributed with the three symmetry-related wave vectors $\bm{Q}_\alpha$. Obviously, this phase is nematic, since neither the VDW nor the SDW order is distributed with all the three symmetry-related wave vectors. The physical picture of this result is as follow. Considering that the three wave vectors $\bm{Q}_\alpha$ are all antiferromagnetic-like, the $z$-component of the SDW order will be most unfavored by the uniform Zeeman term and thus it would be kicked out from the 3D ``easy plane" for the polarization of any DW order; the VDW order parameter is completely blind to the Zeeman coupling and thus \sout{it's}{\color{red}it is} maximized and fully occupies a wave vector; the $x,y$-components of the SDW sit in between the two and occupy the remaining two wave vectors.

The relation between the SO(4) and the SU(2)$_K\times$SU(2)$_{K^\prime}$ symmetries, and the consequent degeneracy between the SDW and VDW orders have been clarified in Refs.~\cite{LiangFu,VishwanathYou} previously. However, the role of the degeneracy among the symmetry-related wave vectors is first thoroughly investigated here. In this work, we reveal that the combination of the two aspects will bring about the TRS-breaking chiral SO(4) spin-valley DW state with intriguing properties, whose energy is reasonably lower than that of the 3Q-VDW state proposed in Ref.~\cite{LiangFu}. Further more, our results are more different from those in Refs.~\cite{LiangFu,VishwanathYou} for the cases of $J_H\ne 0$ (which will be studied in the next section). Briefly, both Refs. \cite{LiangFu} and \cite{VishwanathYou} take the viewpoint that since the SDW and VDW are degenerate at $J_H=0$, one naturally conjectures that for $J_H>0$ ($J_H<0$) the VDW (SDW) will beat the other order. However, \sout{it's}{\color{red}it is} pointed out here that the SDW and VDW generally can be mixed. For $J_H=0$, they are mixed into the chiral SO(4) DW, whose three mutually perpendicular vectorial order parameters can be globally arbitrarily rotated in the $\mathbb{R}^4$ space, forming a degenerate-ground-state set. Then the realistic tiny SU(2)$_K\times$SU(2)$_{K^\prime}$-symmetry-breaking $J_H$ term acts as a perturbation upon this degenerate-ground-state set, whose consequence is to select in this set its favorite states with special polarization directions of the three mutually perpendicular vectorial DW order parameters. As a result, for $J_H\rightarrow0^{-}$ we get pure chiral SDW, while for $J_H\rightarrow0^{+}$ we get a nematic DW state with one stripy VDW component mixed with two SDW components, instead of the pure isotropic VDW suggested by Refs.~\cite{LiangFu,VishwanathYou}. More details of these results will be presented in the next section.

\begin{figure}
\centering
\includegraphics[width=0.49\textwidth]{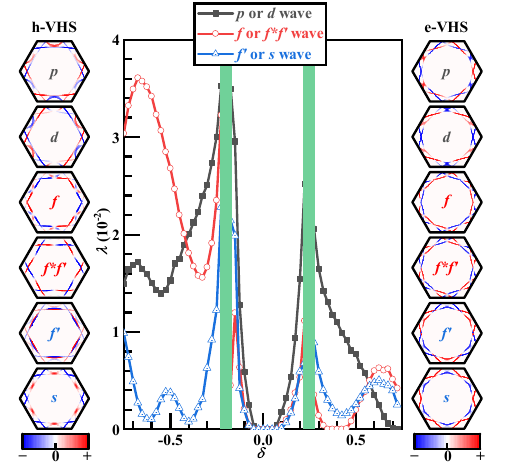}
\caption{The largest pairing eigenvalues $\lambda$ vs doping for all possible pairing symmetries under $U=1.1$ meV. Note about the degeneracy between the $p$- and $d$-, the $f'$- and $s$-, and the $f$- and $f{*}f'$-wave pairings, respectively, referred to Fig.~\ref{symmetry}. The degenerate $p$ and $d$ wave pairings dominate other pairings near the two VH dopings, see the two regimes covered with green rectangles, which represent for the chiral SO(4) DW phase. The insets on both sides show the normalized gap functions for all possible pairing symmetries near the two VH dopings .}\label{SC}
\end{figure}

\begin{figure}[htbp]
\centering
\includegraphics[width=0.48\textwidth]{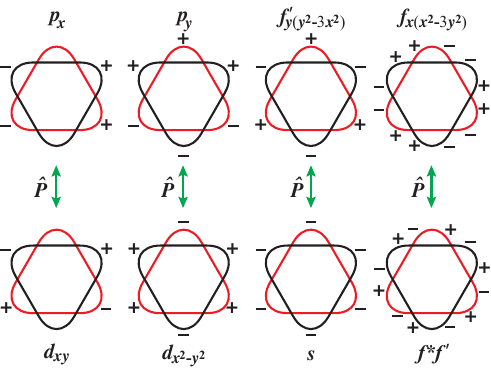}
\caption{One-to-one mapping between the triplet pairings (the first row) and singlet pairings (the second row) under the operation $\hat P$. The red and black curves represent the FSs contributed from the $K$ and $K'$ valleys, respectively.}
\label{symmetry}
\end{figure}

\subsection{Degeneracy between singlet and triplet SCs}

The doping-dependences of the largest pairing eigenvalues for all the pairing symmetries are plotted in Fig.~\ref{SC}, where the gap form factors $\Delta_{\alpha}(\bm{k})$ (determined by Eq.~\eqref{gapeq}) near the two VHS points are shown on both sides. The two green rectangles near the e-VHS and the h-VHS give the regimes for the chiral SO(4) spin-valley DW studied above where $U>U_c^{(s)}=U_c^{(v)}$,  and the remaining regimes support the SC phases. In the regimes near the VHS, the degenerate $p$- and $d$-wave pairings are the leading pairing symmetries, while in the over doped regimes far away from the VHS, the degenerate $f_{x(x^2-3y^2)}$- and $f_{x(x^2-3y^2)}*f^{\prime}_{y(y^2-3x^2)}$- wave pairings become the leading symmetries.

The most remarkable feature of Fig.~\ref{SC} lies in that there is a one-to-one corresponding degeneracy between the triplet and singlet pairings, i.e. the $p$- and $d$-pairing degeneracy, the $f'$- and $s$-pairing degeneracy, and the $f$- and $f{*}f'$-pairing degeneracy, see Fig.~\ref{symmetry}. Similar to the degeneracy between the inter-valley SDW and the VDW, the degeneracy between the inter-valley singlet and triplet pairings originates from that they are related by the unitary symmetry operation $\hat P$ defined in Eq.~\eqref{unitary_symmetry}. Concretely, the following singlet and triplet pairings with order parameters

\begin{subequations}\label{singlet_triplet_pairing}
\begin{align}
\hat {O}^{(s)}_{\rm SC}&=\sum_{mv\bm{k}\in FS}\left[
\hat c_{mv\bm{k}\uparrow}
\hat c_{m\bar{v} \bar {\bm k}\downarrow}
- \hat c_{mv\bm{k}\downarrow}
\hat c_{m\bar v\bar {\bm k} \uparrow}
\right]\Delta_{mv}(\bm{k}),\\
\hat {O}^{(t)}_{\rm SC}&=-\sum_{mv\bm{k}\in FS}\left[
\hat c_{mv\bm{k}\uparrow}
\hat c_{m\bar{v}\bar {\bm {k}}\downarrow}
+ \hat c_{mv\bm{k}\downarrow}
\hat c_{m\bar{v} \bar {\bm k}\uparrow}
\right] v\Delta_{mv}(\bm{k}),
\end{align}
\end{subequations}
are related as
\begin{equation}\label{singlet_triplet_relating}
\hat P^{\dagger}\hat{O}^{(s)}_{\rm SC}\hat{P}=\hat{O} ^{(t)}_{\rm SC},
\end{equation}
where $\bm \bar{\bm k}\equiv-\bm k, \bar{v}\equiv-v$ and the operator $\hat{P}$ is defined by Eq. (\ref{unitary_symmetry}). Note that in the weak-pairing limit only the electrons on the FS participate in the pairing, and an electron state on the $(mv)$-th band with momentum $\bm{k}$ can only pair with its TR-partner, i.e. the state on the $(m\bar{v})$-th band with momentum $\bar{\bm{k}}$. The condition $mv\bm{k}\in FS$ defines $v$ as an implicit function of $\bm{k}$, and from Fig.~\ref{symmetry} we have $v_{\bm \bar{\bm k}}=-v_{\bm{k}}$, suggesting that  $v_{\bm{k}}$ is an odd function of $\bm{k}$.  Equations \eqref{singlet_triplet_pairing} and \eqref{singlet_triplet_relating} suggest that a singlet pairing with even-parity gap function $\Delta_{mv}(\bm{k})$ can be mapped to a triplet pairing with odd-parity gap function $-v_{\bm{k}}\Delta_{mv}(\bm{k})$. In Fig.~\ref{symmetry}, the distributions of the gap signs for all possible pairing symmetries are schematically shown, where the listed one-to-one mapping between different singlet and triplet pairings can well explain the singlet-triplet degeneracy shown in Fig.~\ref{SC}.

Similar to the degeneracy between the SDW and VDW orders, the degeneracy between the singlet and triplet SCs also originates from the SU(2)$_K\times$SU(2)$_{K^\prime}$ symmetries. However, there is an important difference between them: for the SC, there is only one ``nesting vector" or ``wave vector", i.e., $\bm{Q}=0$ in the particle-particle channel, which is the center-of-mass momentum of a Cooper pair. As a result, the singlet-triplet degeneracy for SC is always tenable, leading to degenerate ground-state energies for singlet and triplet SCs and hence their arbitrary mixing. Such a degeneracy can only be lifted up by adding a weak inter-valley Hund's-rule coupling that will be studied in the next section.
\begin{figure}[htbp]
\centering
\includegraphics[width=0.48\textwidth]{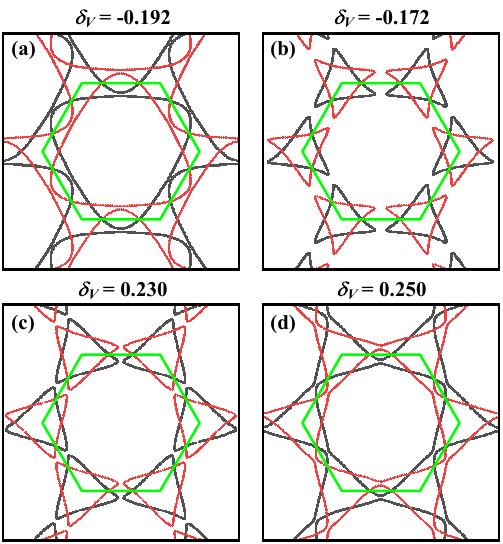}
\caption{FSs on the over and under doping sides of the h-VHS point (a, b) and e-VHS point (c, d) with the same filling deviation of 0.01. The FSs show better nesting behavior on the over doping side than on the under doping side. Other denotations and parameters are the same with those in Fig.~\ref{band}.}\label{FSs}
\end{figure}

\begin{figure}[htbp]
\centering
\includegraphics[width=0.48\textwidth]{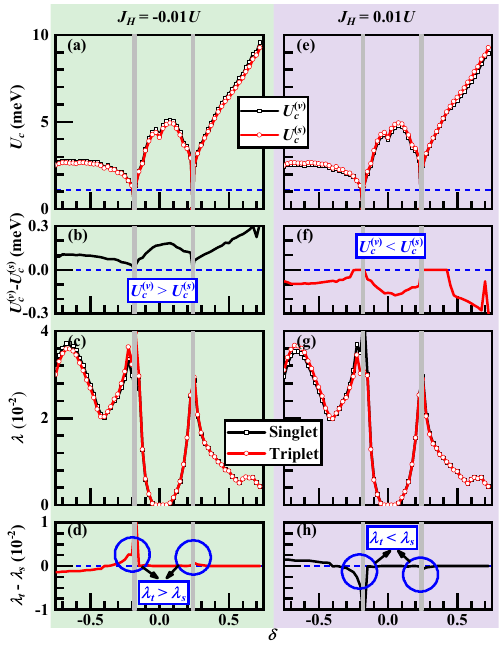}
\caption{Doping dependences of $U^{(s)}_c$, $U^{(v)}_c$ (a, e) and their difference (b, f) and of the largest eigenvalues $\lambda$ for the singlet-pairing, the triplet pairing (c, g) and their difference (d, h) with $J_H=-0.01U$ for the left column and $J_H=0.01U$ for the right column. In the calculations for (c),(d),(g) and (h), $U=1.1$meV is adopted.}
\label{JHUc}
\end{figure}

The doping-dependence of the superconducting $T_c$ shown in Fig.~\ref{SC} exhibits two asymmetric behaviors consistent with experiments. One is the asymmetry with respect to the CNP: the $T_c$ at the negative dopings is much higher than that at the positive dopings, which is due to the higher DOS for the former case than that for the latter case (see Fig.~\ref{band}(b)). Such an asymmetric behavior is well consistent with both the experiments of Y. Cao, et al, in Ref.~\cite{caoyuan1} and the observations of  M. Yankowitz, et al, in Ref.~\cite{Yankowitz}. The other asymmetry is with respect to each VH doping: the $T_c$ on the higher-doping side of each VH point is higher than that on its lower-doping side. This asymmetry is attributed to the asymmetric situations of the FS-nesting on the two sides of each VH doping, see Fig.~\ref{FSs} which indicates that the FSs are better nested at the higher-doping side of each VH doping than those at its lower-doping side. As a result, the susceptibility and hence the effective pairing interaction on the higher-doping side of each VH doping are stronger than those on the other side, leading to the higher $T_c$ on the higher-doping side. This asymmetric behavior is also well consistent with both experiments in Refs.~\cite{caoyuan1} and \cite{Yankowitz}. The consistence of these two asymmetric doping-dependent behaviors of the $T_c$ with the experiments suggests that the SC pairing mechanism in the MA-TBG should be consistent with that we proposed, i.e. exchanging the spin-valley DW fluctuations.

\section{Results with weak inter-valley exchange interactions ($J_H\neq0$)}

For the realistic material of the MA-TBG, theoretical analysis suggests that there exists a very weak inter-valley Hund's -rule exchange interaction with strength $J_H\approx 0.01U$ \cite{VishwanathYou,LiangFu,JYLee} which has been neglected in Sec.~\ref{SecWithoutJH}. As in the case of $J_H=0$, the SU(2)$_K\times$SU(2)$_{K^\prime}$ symmetry brings about the SDW-VDW degeneracy at the critical point and the singlet-triplet degeneracy for SCs, \sout{it's}{\color{red}it is} necessary to add the tiny symmetry-breaking $J_H$-term to lift up these degeneracies. Further more, this symmetry also leads to the chiral SO(4) spin-valley DW ground state which hosts three vectorial DW order parameters, whose polarization directions need to be pinned down by the tiny symmetry-breaking $J_H$ term. In this section, we focus on the infinitesimal $J_H$ term, including $J_H\to 0^{-}$ and $J_H\to 0^{+}$, and investigate its influence on the ground state of the MA-TBG.  The two cases will be studied separately in the following.

\subsection{$J_H\to 0^{-}$}

For the case of $J_H\to 0^{-}$, we set $J_H=-0.01U$ and redo the RPA calculations. The results of our RPA calculations are shown in Figs.~\ref{JHUc}(a) to \ref{JHUc}(d). The doping-dependence of the critical interaction strength $U_{c}^{(s,v)}$ shown in Figs.~\ref{JHUc}(a) suggests $U_{c}^{(v)}>U_{c}^{(s)}$, as is verified by the broadened $U_{c}^{(v)}-U_{c}^{(s)}>0$ shown in Figs.~\ref{JHUc}(b). This result suggests that a negative $J_H$ favors the SDW order. In such a case, we redo the energy optimization of the Hamiltonian \eqref{H_Hubbard} in the mixed spin-valley DW state determined by Eq.~\eqref{HMF}, with the same variational parameters. Our result reveals that the pure chiral SDW states \cite{Fanyang} obtained in Sec.~\ref{ConsequenceDegeneracy} are the ground states. The physical picture for the evolution from the chiral SO(4) spin-valley DW in the case of $J_H=0$ to the chiral SO(3) SDW state in the case of $J_H\to 0^{-}$ is simple: in the former case, due to the SO(4) symmetry, the four axes for each spin-valley DW vectorial order are equally favored, which leads to the free rotation of that vectorial order in the $\mathbb{R}^4$ space; however, in the latter case, the VDW-axis for each DW order parameter is disfavored and the left three SDW-axes form the $\mathbb{R}^3$ easy ``plane", within which the SDW vectorial orders can arbitrarily rotate.

\begin{figure*}[htbp]
\centering
\includegraphics[width=0.9\textwidth]{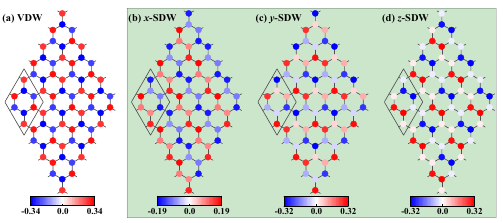}
\caption{The real-space distributions of the scalar inter-valley charge density (a) and the three components of the vectorial inter-valley spin density (b-d) for a typical ground state configuration with $\bm\varDelta_1=(0.020, 0.41, 0.32, 0.51)$ meV, $\bm \varDelta_2=(0.72, -0.0019, 0.08, 0.0)$ meV, and $\bm \varDelta_3=(-0.077, 0.13, 0.55, -0.44)$ meV in the nematic DW phase for $J_H=0.01U$. Note that the pattern in (a) nearly takes only one wave vector, i.e. $\bm{Q}_2$, while those in (b-d) take both $\bm{Q}_1$ and $\bm{Q}_3$.}
\label{SO4DWJH}
\end{figure*}

The chiral SDW state obtained here has similar properties in many aspects with the same phase obtained previously in other contexts \cite{TaoLi,Martin,Kato,Ying,Fanyang}. The real-space configuration of the chiral SDW state also has four sublattices. This ground state hosts four branches of gapless Goldstone modes, including three spin-wave modes brought about by the spin-SU(2) symmetry breaking and one extra valley-wave modes caused by the valley-U(1) symmetry breaking. At finite temperature, the gapless Goldstone-mode fluctuations will also destroy the long-range DW order, leaving short-ranged DW fluctuations with long correlation length below some characteristic temperature. Further more, the TRS breaking of this state can survive finite temperature. The topological properties of this state can also be nontrivial with nonzero Chern number, as long as an SDW gap opens at the Fermi level.

However, the close proximity of the chiral SDW state obtained here for $J_H\to 0^{-}$ to the chiral SO(4) spin-valley DW state for $J_H=0$ makes it different from those in other contexts \cite{TaoLi,Martin,Kato,Ying,Fanyang} in the aspect of the response to a weak magnetic field. The condition $J_H\to 0^{-}$ and the applied weak magnetic field studied in the Sec.~\ref{SCDegeneracy} both have the effect of pinning down the directions of the polarizations of the DW orders. However, the effects brought about by them conflict: while the former case disfavors the VDW, the latter favors it. Considering that the $J_H$ in real materials is very weak, a weak magnetic field (a few Tesla) is enough to overcome its effects. As a result, the weak applied magnetic field would drive the isotropic chiral SDW state here into a nematic DW state containing one nematic VDW order and two nematic SDW orders. Such an effect can be easily checked by experiments.

The doping-dependence of the largest pairing eigenvalues for the singlet and triplet pairing symmetries is shown in Fig.~~\ref{JHUc}(c). Clearly the tiny SU(2)$_K\times$SU(2)$_{K^\prime}$-symmetry-breaking $J_H$-term leads to the split between the singlet and triplet pairings. Concretely, near the VHS the triplet $p$-wave pairing wins over the singlet $d$-wave one and becomes the leading pairing symmetry, while far away from the VHS in the over doped regime the singlet $f_{x(x^2-3y^2)}*f^{\prime}_{y(y^2-3x^2)}$-wave pairing beats the triplet $f_{x(x^2-3y^2)}$- wave pairing and serves as the leading pairing symmetry. In the experiments reported in Refs.~\cite{caoyuan1} and \cite{Yankowitz}, the SC is mainly detected near the VHS. Therefore, the experiment-relevant pairing symmetry in the case of $J_H\to 0^{-}$ should be triplet $p$-wave pairing. As the $p$-wave belongs to the 2D irreducible representation, the degenerate $p_x$- and $p_y$-wave pairings would always be mixed into the $p_x\pm ip_y$ form to lower the ground-state energy, i.e. the $p+ip$ for abbreviation, as verified by our numerical results. This state is topologically nontrivial. As the $J_H$ is very weak, the two asymmetric behaviors of the doping-dependence of the superconducting $T_c$ shown in Fig.~\ref{JHUc}(c) are similar with the case of $J_H=0$ shown in Fig.~\ref{SC}, which are consistent with experiments.

\subsection{$J_H\to 0^{+}$}

The RPA results for $J_H\to 0^{+}$ are shown in Figs.~\ref{JHUc}(e)- \ref{JHUc}(h). Figures~\ref{JHUc}(e) and \ref{JHUc}(f) obviously show $U_c^{(s)}>U_c^{(v)}$, suggesting that the VDW is more favored than the SDW here. However, this does not mean that the ground state for general realistic $U>U_c^{(s)}\approx U_c^{(v)}$ is in the pure VDW phase, due to the following reason. The tiny positive $J_H$ term as a perturbation on the chiral SO(4) DW state, its only role is to set the VDW-axis as an easy axis for the three vectorial DW order parameters $\bm\varDelta_{\alpha}$ to orient in the $\mathbb{R}^{4}$ space. However, among the three mutually perpendicular $\bm\varDelta_{\alpha}\ (\alpha=1,2,3)$, at most one lucky $\bm\varDelta_{\alpha}$ is given the opportunity to orient toward the VDW-axis, with the remaining two still residing in the $\mathbb{R}^{3}$ SDW-``plane", leading to a mixed VDW and SDW ordered state. Such an argument is consistent with the following numerical results for the succeeding MF-energy minimization.  Firstly, the three relative phase angles between the VDW and SDW orders are $\theta_\alpha\approx\frac{\pi}{2}$, keeping the approximate SO(4) symmetry. Secondly, among the three DW order parameters $\bm\varDelta_\alpha$, an arbitrarily chosen one, say $\bm\varDelta_2$, takes the form of $\bm\varDelta_2\approx(\Delta, 0,0,0)$, while the remaining two i.e. $\bm\varDelta_1$ and $\bm\varDelta_3$, both take the form of $(0,\Delta_1,\Delta_2,\Delta_3)$ with $\bm\varDelta_1\perp\bm\varDelta_3$.  This result suggests that for $J_H\to 0^{+}$, we obtain a spin-valley DW ordered ground state with one scalar VDW order parameter accompanied by another two mutually perpendicular vectorial SDW order parameters, with the three DW order parameters randomly distributed with the three symmetry-related wave vectors $\bm{Q}_\alpha$.

In Fig.~\ref{SO4DWJH}, the real-space distributions of the inter-valley charge and spin densities defined in Eq.~\eqref{VDW_SDW_ORDER} are shown for a typically chosen group of DW order parameters for this phase, i.e. $\bm\varDelta_1=(0.020,\ 0.41,\ 0.32,\ 0.51)$, $\bm \varDelta_2=(0.72,\ -0.0019,\ 0.08,\ 0.0)$, and $\bm \varDelta_3=(-0.077,\ 0.13,\ 0.55,\ -0.44)$. As the VDW order in this DW state nearly only takes one wave vector $\bm{Q}_2$ among the three symmetry-related ones $\{\bm{Q}_\alpha\ (\alpha=1,2,3)\}$, the inter-valley charge density shown in Fig.~\ref{SO4DWJH}(a) exhibits a nematic stripy structure, which spontaneously breaks the $C_3$ rotational symmetry of the original lattice. Note that the extension direction of the charge stripe can be arbitrary among the three symmetry-related directions. Such a nematic stripy distribution of the inter-valley charge density is related to the recent STM experiments \cite{Jiang, Kerelsky}. Note that the $C_3$-symmetry breaking here for the inter-valley charge density can be delivered to the intra-valley one relevant to the STM based on the Ginsberg-Landau theory, as it cannot be excluded that the two orders are coupled. Here we have provided a simple understanding toward these experimental observations based on the spontaneous breaking of the $C_3$ symmetry, which suggests that the $J_H\to 0^{+}$ is more realistic for the MA-TBG. It's interesting that the ground state of the system is not a pure nematic VDW, but it also comprises of two additional nematic SDW orders with equal amplitudes, as shown in Fig.~\ref{SO4DWJH}(b-d) for the three components of the inter-valley spin density. Here we propose that a spin-dependent STM can detect such a nematic spin order, which coexists with the already-detected nematic stripy charge order.

This spin-valley DW ground state hosts four branches of gapless Goldstone modes, including three spin-wave modes brought about by the spin-SU(2) symmetry breaking and one extra valley-wave modes caused by the valley-U(1) symmetry breaking. At finite temperature, the DW fluctuations will also destroy the long-range DW order, leaving short-ranged DW fluctuations with long correlation length below some characteristic temperature. However, the VDW order parameter, the TRS breaking, and the $C_3$-symmetry breaking can survive the finite temperature, as they are discrete symmetry breakings. Besides, the topological properties of this state can also be nontrivial if \sout{it's}{\color{red}it is} insulating. Therefore, at finite temperature for $J_H\to 0^{+}$, we obtain a nematic VDW state with TRS breaking, which simultaneously hosts strong SDW fluctuations with long spin-spin correlation length.

The doping-dependence of the largest pairing eigenvalues for the singlet and triplet pairing symmetries are shown in Fig.~~\ref{JHUc}(g) for $J_H\to 0^{+}$. Consequently, near the VHS the singlet $d$-wave pairing wins over the triplet $p$-wave pairing and becomes the leading pairing symmetry, while far away from the VHS in the over doped regime the triplet $f_{x(x^2-3y^2)}$-wave pairing beats the singlet $f_{x(x^2-3y^2)}*f^{\prime}_{y(y^2-3x^2)}$-wave pairing and serves as the leading pairing symmetry. The experiment-relevant pairing symmetry near the VH dopings in this case should be singlet $d$-wave pairing, which takes the form of topological $d+id$ pairing state.  As the $J_H$ is very weak, the two asymmetric behaviors of the doping-dependence of the superconducting $T_c$ shown in Fig.~\ref{JHUc}\add{(g)} are also clear, which are consistent with experiments.

\section{Conclusion and Discussion}
In conclusion, by adopting realistic band structure and interactions, we have performed a thorough investigation on the electron instabilities of the MA-TBG driven by FS-nesting near the VH dopings. A particular attention is paid to the approximate SU(2)$_K\times$SU(2)$_{K^\prime}$ symmetry and the three-folded wave-vector degeneracy brought about by the $D_3$-rotational symmetry of the system. At the SU(2)$_K\times$SU(2)$_{K^\prime}$-symmetric point with $J_H=0$, we obtain the chiral SO(4) spin-valley DW state. This state is a generalization of the 3Q chiral SDW state to the $\mathbb{R}^4$ VDW-SDW order-parameter space, which is a novel state possessing a series of exotic properties. The leading pairing symmetries are degenerate singlet $d+id$ and triplet $p+ip$. For $J_H\to 0^{-}$, we obtain the pure 3Q chiral SDW state, and triplet $p+ip$-wave pairing. For $J_H\to 0^{+}$, we obtain a nematic DW state with mixed SDW and stripy VDW orders, and singlet $d+id$-wave pairing. The stripy inter-valley charge-density pattern in this nematic state is consistent with recent STM experiments, suggesting that $J_H\to 0^{+}$ is more realistic for the MA-TBG. These results are summarized in Fig.~\ref{classification}. Besides, the two asymmetric doping-dependent behaviors of the pairing phase diagram shown in Fig.~\ref{SC} and \ref{JHUc} are well consistent with experiments, suggesting the relevance of the exchanging-DW-fluctuations pairing mechanism for the MA-TBG.

The $p_{x,y}$-orbital TB model on the honeycomb lattice adopted here is criticized to be topologically problematic \cite{Zou2018, Po2} for the CNP. However, here we focus on the doped case, particularly on the VHS, and therefore only the low-energy band structure near the FS will matter. For more accurate band structure, we can adopt the continuum-theory band structure directly \cite{MacDonald}, which is not only complicated but also has the difficulty of how to properly put in the interaction terms. Alternatively, later than the present work, part of the present authors have recently adopt the faithful TB model \cite{Po2} with five bands per valley per spin which can properly deal with the band topology to study the problem. Although the band structure of that model is much more complicated than that of our present model, the results published in Ref.~\cite{Zhang2020dw} are qualitatively consistent with those obtained in this work.  The reason lies in that the physics discussed in this paper only relies on the approximate SU(2)$_K\times$SU(2)$_{K^\prime}$ symmetry, the valley-U(1) symmetry and the presence of three-fold degenerate nesting vectors which originate from the $D_3$-rotational symmetry of the material. These symmetries do not depend on the details of the band structures.

Note that the nesting vectors $\bm{Q}_{\alpha}$ in our model only locate along the $\varGamma_M M_M$ lines, but not exactly at the $M_M$ points. If we adopt the accurate value of $\bm{Q}_{\alpha}$ (generally incommensurate) to build our VDW or SDW order parameters, the unit cell would be huge or even infinite, which brings great difficulty to the calculations. Further more, the relation $\bm{Q}_\alpha\ne -\bm{Q}_\alpha$ might bring further difficulty to the calculations. However, as the main physics revealed here only relies on the three-folded wave-vector degeneracy brought about by the $D_3$ symmetry of the system, we argue that the accurate values of $\bm{Q}_{\alpha}$ should not influence the main results.

The chiral character of the SO(4) DW state predicted in this work is lack of experiment evidence presently. The reason for this might lie as follow. This state is formed as a consequence of the competition among the three degenerate wave vectors caused by the three-fold rotation symmetry of the system. In realistic system, there might be such factors as the strain which will break the exact three-fold rotation symmetry. As a result, only one of the three wave vectors might win and be realized, which breaks the chiral DW state. Therefore, the state obtained in our work needs ideal experimental condition to be realized, which might be realized in the future. It's also possible that the weak-coupling start point, as well as the concrete formula of the multi-orbital Hubbard interactions adopted here does not apply to the real material of the magic-angle twisted bilayer graphene system. However, the physics revealed here might apply to other systems with similar degrees of freedom.

\section*{Acknowledgements}
 We are grateful to the stimulating discussions with Noah Fan-Qi Yuan, Yi-Zhuang You, Jun-Wei Liu, Xi Dai, and Long Zhang. This work is supported by the National Natural Science Foundation of China under the Grants No.12074031 (F. Y.), No. 12074037 (Y.Z.), No. 11922401 (C.-C.L.), No. 11874292, No. 11729402 and No. 11574238(Y.W.), No. 11861161001 and No. 12141402 (W.-Q.C.).Z.-C.G. is supported by funding from Hong Kong¡¯s Research Grants Council (NSFC/RGC Joint Research Scheme No. N-CUHK427/18 and General Research Fund Grant No. 14302021). W.-Q.C. is supported by the Science,Technology and Innovation Commission of Shenzhen Municipality (No. ZDSYS20190902092905285), Guangdong Basic and Applied Basic Research Foundation under Grant No. 2020B1515120100, Shenzhen-Hong Kong Cooperation Zone for Technology and Innovation (Grant No. HZQBKCZYB-2020050), and Center for Computational Science and Engineering at Southern University of Science and Technology.

\appendix

\section{Tight-binding Hamiltonian $H_{\rm TB}$}\label{appendixHTB}

This Appendix provides some details for the TB Hamiltonian $H_{\rm TB}$ in Eq.~\eqref{HTB}, including its connection with the Slater-Koster formalism and the U(1)-valley symmetry. In addition, how to transform it from the $p_{x,y}$-orbital representation to the valley representation is shown.

The proposed simplest TB model for the MA-TBG possesses two orbitals of $p_x$ and $p_y$ on each lattice site \cite{Yuan2018, Po2018, Kang1, Fanyang}, holding the form,
\begin{align}\label{HU0}
\hat H_{0}=\sum_{j\mu,j'\mu'\sigma} t_{j\mu,j'\mu'}\hat c^{\dagger}_{j\mu\sigma}\hat c_{j'\mu'\sigma}-\mu_c \sum _{j\mu\sigma}\hat c^{\dagger}_{j\mu\sigma}\hat c_{j\mu\sigma},
\end{align}
where $\hat c_{j\mu\sigma}$ is the annihilation operator of the electron with the $\mu$-th ($\mu=x,\ y$ represents $p_x$ or $p_y$) orbital and spin $\sigma$ on the $j$-th site. $\mu_c$ is the chemical potential and $t_{j\mu,j'\mu'}$ is the hopping integral between the $\mu$ and $\mu'$ orbitals on the $j$th and $j'$th sites, respectively.

In the case with $D_6$ symmetry, the hopping integral can be constructed \cite{Fanyang} via the Slater-Koster formalism \cite{JMCM} based on the coexisting $\sigma$ and $\pi$ bondings \cite{DMSarma, CMWu,GMF, CMCMLiu, FMYang}, namely,
\begin{equation}\label{slater_koster}
t_{j\mu,j'\mu'}=t_{\sigma}^{jj'}\cos\theta_{\mu,jj'}\cos\theta_{\mu',jj'}+t_{\pi}^{jj'}\sin\theta_{\mu,jj'}\sin\theta_{\mu',jj'},
\end{equation}
with $\theta_{\mu,jj'}$ denotes the angle from the direction of $\mu$ to that of $\bm{r}_{j'}-\bm{r}_{j}$. The Slater-Koster parameters of $t_{\sigma}^{jj'}$ and $t_{\pi}^{jj'}$ represent the parts of the hopping integrals caused by $\sigma$ and $\pi$ bonds between the $j$th and $j'$th sites, respectively.

To reflect the U(1)-valley symmetry, the above Slater-Koster Hamiltonian \eqref{HU0} can be transformed into the valley representation via $\hat c_{j\pm\sigma}=(\hat c_{jx\sigma}\pm i\hat c_{jy\sigma})/\sqrt{2}$ with $\pm$ representing the $K$ and $K'$ valley. As required by the U(1)-valley symmetry, the inter-valley hopping terms should vanish, which leads to,
\begin{subequations}
\begin{align}\label{tij}
2t_{\sigma}^{jj'}\cos\theta_{x,jj'}\cos\theta_{y,jj'}+2t_{\pi}^{jj'}\sin\theta_{x,jj'}\sin\theta_{y,jj'}&=0,\\
t_{\sigma}^{jj'}(\cos^2\theta_{x,jj'}{-}\cos^2\theta_{y,jj'})+t_{\pi}^{jj'}(\sin^2\theta_{x,jj'}{-}\sin^2\theta_{y,jj'})&=0.
\end{align}
\end{subequations}
Since $\theta_{y,jj'}=\theta_{x,jj'}-\frac{\pi}{2}$, we get
\begin{align}\label{tiij}
t&_{\sigma}^{jj'}=t_{\pi}^{jj'}\equiv t^{jj'}.
\end{align}
Substituting Eq.~\eqref{tiij} into Eq.~\eqref{slater_koster}, we have,
\begin{align}
t_{j\mu,j'\mu'}=t^{jj'}\delta_{\mu\mu'}.
\end{align}
Up to the third neighbor hoppings, the Hamiltonian \eqref{HU0} turns into \cite{Yuan2018, Liang1Fu2},
\begin{align}\label{H0}
\hat H_0=\sum_{\alpha=1}^3\sum _{ \langle jj'\rangle_\alpha v \sigma} t_\alpha \left(\hat c^{\dagger}_{jv\sigma}\hat c _{j'v\sigma}+ {\rm h.c.} \right)-\mu_c\sum _{jv\sigma}\hat c^{\dagger}_{jv\sigma}\hat c_{jv\sigma},
\end{align}
where $v=\pm$ and $\langle jj'\rangle_\alpha$ denotes the $\alpha$th neighboring bond with the hopping strength of $t_\alpha$. This Hamiltonian has the valley-SU(2) symmetry.

For the real material of the MA-TBG, the point-group is $D_3$ instead of $D_6$. The breaking of $D_6$ down to $D_3$ brings about the Kane-Mele type of valley-orbital coupling, i.e.,
\begin{align}\label{H1}
\hat H_1=&\sum_{\alpha=1}^3\sum _{ \langle jj'\rangle_\alpha\sigma} t'_\alpha\left[(\hat c^{\dagger}_{j\sigma}\times \hat c _{j'\sigma} )_z+{\rm h.c.}\right]\nonumber\\
=&-i\sum_{\alpha=1}^3\sum _{ \langle jj'\rangle_\alpha\sigma}t'_\alpha\left(\hat c^{\dagger}_{j+\sigma}\hat c _{j'+\sigma}-\hat c^{\dagger}_{j-\sigma}\hat c_{j'-\sigma} \right)+{\rm h.c.},
\end{align}
where $\hat c_{j\sigma}=(\hat c_{jx\sigma},\ \hat c_{jy\sigma})^T$ and $t'_\alpha$ describes the $\alpha$th neighboring coupling strength.

Combining $\hat H_0$ and $\hat H_1$, we arrive at the TB Hamiltonian expressed in the Eq.~\eqref{HTB} of the main text, which satisfies the U(1)-valley symmetry \cite{Yuan2018, Liang1Fu2}.

\section{More information on RPA approach}
\label{RPAapproach}

In this appendix, we provide the detailed informations on the RPA approach, including the explicit form of the non-interaction susceptibility $\chi^{(0)}$, the interaction matrices $\tilde U^{(s)}$ and $\tilde U^{(c)}$, and the effective pairing interaction vertex $V^{\alpha\beta}(\bm k, \bm k')$.

The form of $\chi^{(0)}$ is given by
\begin{align}
\chi^{(0)l_1l_2}_{l_3l_4}\left(\bm{q},i\omega\right)=&\frac{1}{N}\sum_{\bm k, \alpha\beta}{n_F(\tilde\varepsilon_{\bm k+\bm q}^\beta)-n_F(\tilde\varepsilon_{\bm k}^\alpha)\over \tilde\varepsilon_{\bm k}^\alpha-\tilde\varepsilon_{\bm k+\bm q}^\beta+i\omega} \nonumber\\
&\times\xi_{l_1}^{\alpha*}(\bm k)\xi_{l_2}^\beta(\bm k{+}\bm q)\xi_{l_4}^{\beta*}(\bm k+\bm q)\xi_{l_3}^\alpha(\bm k),
\end{align}
where $n_F(\tilde\varepsilon_{\bm k}^\alpha)$ is the Fermi-Dirac distribution. $\alpha$ and $\beta$ represent the the combined index $(mv)$ in Eq.~\eqref{HTB}. $\tilde\varepsilon_{\bm k}^\alpha$ and $\xi^\alpha(\bm k)$ are the energy level and corresponding eigenstate at the wave vector $\bm k$ for the $\alpha$-th band, both of which are determined by Eq.~\eqref{HTB}. In the RPA level, the renormalized spin and charge susceptibilities have been given in Eqs.~\eqref{Renorm_SUS_spin} and \eqref{Renorm_SUS_charge}, in which
\begin{subequations}
\begin{align}
&\tilde{U}^{(s)}=U^{(s)}-2S,\\
&\tilde{U}^{(c)}=U^{(c)}+2S.
\end{align}
\end{subequations}
Labelling orbitals $\left \{ p^{A}_{+},p^{A}_{-},p^{B}_{+},p^{B}_{-}, \right \}$ as $\left \{ 1,2,3,4 \right \}$, the explicit forms of $U^{(s)}$, $U^{(c)}$, and $S$ are given as follow. Firstly, the nonzero elements of $U^{(s)l_1l_2}_{l_3l_4}$ are:
\begin{subequations}
\begin{align}
&U^{(s)11}_{11}=U^{(s)22}_{22}=U^{(s)33}_{33}=U^{(s)44}_{44}=U,\\
&U^{(s)11}_{22}=U^{(s)22}_{11}=U^{(s)33}_{44}=U^{(s)44}_{33}=-2J_{H},\\
&U^{(s)12}_{12}=U^{(s)21}_{21}=U^{(s)34}_{34}=U^{(s)43}_{43}=U.
\end{align}
\end{subequations}
Secondly, the nonzero elements of $U^{(c)l_1l_2}_{l_3l_4}$ are:
\begin{subequations}
\begin{align}
U^{(c)11}_{11}&=U^{(c)22}_{22}=U^{(c)33}_{33}=U^{(c)44}_{44}=U\nonumber \\
&+4W_2\left[\cos q_1 +\cos q_2 +\cos(q_1 -q_2)\right],                    \\
U^{(c)11}_{22}&=U^{(c)22}_{11}=U^{(c)33}_{44}=U^{(c)44}_{33}=2U+2J_{H}\nonumber\\
&+4W_2\left[\cos q_1 +\cos q_2 +\cos (q_1 -q_2)\right],    \\
U^{(c)12}_{12}&=U^{(c)21}_{21}=U^{(c)34}_{34}=U^{(c)43}_{43}=-4J_{H}-U,\\
U^{(c)11}_{33}&=U^{(c)11}_{44}=U^{(c)22}_{33}=U^{(c)22}_{44}=2W_1\left(1+e^{iq_1}+e^{iq_2}\right) \nonumber \\
&+2W_3\left[2\cos(q_1 -q_2)+e^{i(q_1 +q_2)}\right],\\
U^{(c)33}_{11}&=U^{(c)44}_{11}=U^{(c)33}_{22}=U^{(c)44}_{22}=2W_1\left(1+e^{-iq_1}+e^{-iq_2}\right) \nonumber\\
&+2W_3\left[2\cos(q_1 -q_2)+e^{-i(q_1 +q_2)}\right].
\end{align}
\end{subequations}
Finally, the nonzero elements of $S^{l_1l_2}_{l_3l_4}$ read:
\begin{subequations}
\begin{align}
&S^{11}_{33}=S^{12}_{43}=S^{21}_{34}=S^{22}_{44}=-\frac{J}{2}\left(1+e^{iq_1}+e^{iq_2}\right),\\
&S^{33}_{11}=S^{43}_{12}=S^{34}_{21}=S^{44}_{22}=-\frac{J}{2}\left(1+e^{-iq_1}+e^{-iq_2}\right).
\end{align}
\end{subequations}
In the expressions of $U^{(c)}$ and $S$, $q_{1,2}\equiv \bm{q}\cdot \bm{a}_{1,2}$, where $\bm{a}_{1,2}$ are the two unit vectors of the Moir\'e lattice.

\begin{figure}
\centering
\includegraphics[width=0.48\textwidth]{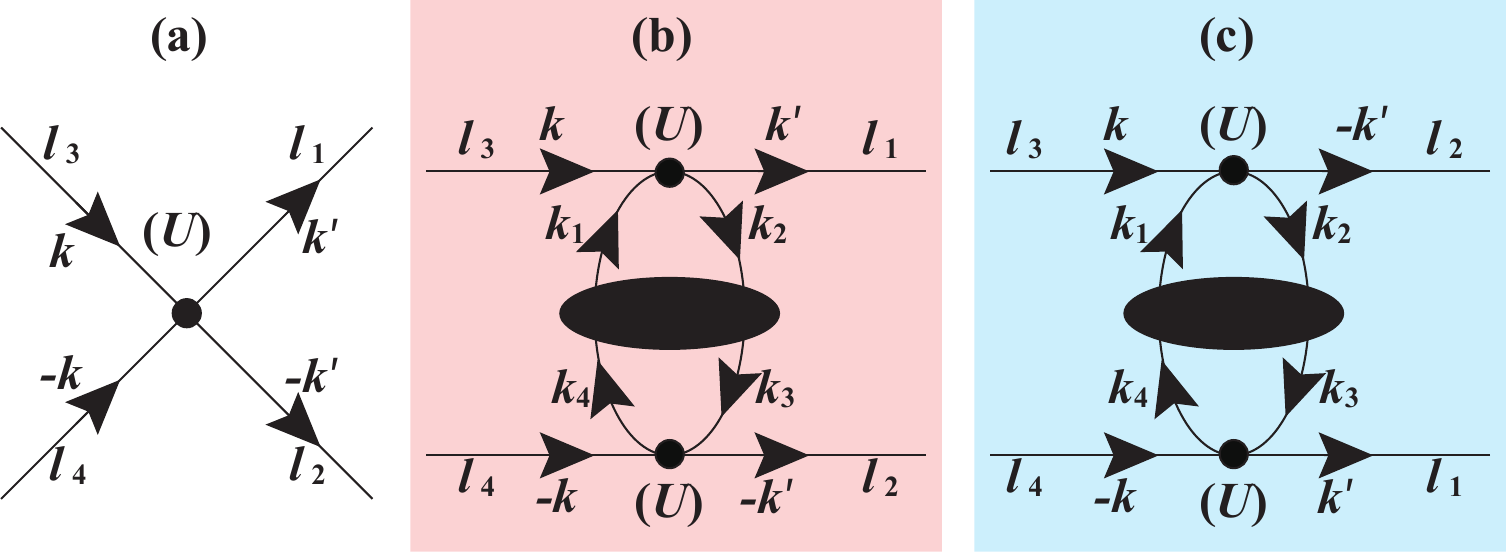}
\caption{Three processes that have contributions to the renormalized effective vertex in the RPA: (a) bare interaction vertex and (b, c) two second order perturbative processes during which spin or charge fluctuations are exchanged between a cooper pair.} \label{RPA_effective}
\end{figure}

In the RPA level, the Cooper pair with momentum and orbital of $(\bm{k}l_3,-\bm{k}l_4)$ could be scattered into $(\bm{k}'l_1,-\bm{k'}l_2)$ by exchanging charge or spin fluctuations, see Fig.~\ref{RPA_effective} which is up to the second order perturbation. These processes induce the following effective interaction,
\begin{align}
V_{\rm eff}=
\frac{1}{N}\sum_{\alpha\beta,\bm{k}\bm{k'}}V^{\alpha\beta}(\bm{k,k'})
\hat c_{\alpha\bm{k}}^{\dagger}
\hat c_{\bar{\alpha}\bm{\bar\bm{k}}}^{\dagger}
\hat c_{\bar{\beta}\bm{\bar\bm{k}}'}
\hat c_{\beta\bm{k}'},\label{pairing_interaction}
\end{align}
where $\bar{\alpha}$ and $\bar{\beta}$ denote the opposite-valley bands of the $\alpha$th and $\beta$th ones, respectively, and $\bm{\bar\bm{k}} = \bm k$. The effective pairing interaction vertex $V^{\alpha\beta}(\bm{k,k'})$ has the form,
\begin{align}
V^{\alpha\beta}(\bm{k,k'})=\!\!\sum_{l_1l_2l_3l_4}\!\!\Gamma^{l_1l_2}_{l_3l_4}(\bm{k,k'},0)\xi_{l_1}^{\alpha,*}(\bm{k})
\xi_{l_2}^{\bar{\alpha},*}({-}\bm{k})\xi_{l_4}^{\bar{\beta}}({-}\bm{k'})\xi_{l_3}^{\beta}(\bm{k'}).\label{effective_potential}
\end{align}
The three processes that have contributions to $\Gamma^{l_1l_2}_{l_4l_3}(\bm{k},\bm{k}')$ are presented in Fig.~\ref{RPA_effective} where (a) denotes the bare interaction vertex and (b, c) represent two second order perturbation processes. During them the spin or charge fluctuations are exchanged within a cooper pair. The effective vertex $\Gamma^{l_1l_2}_{l_3l_4}(\bm{k},\bm{k}')$ is,
\begin{align}
&\Gamma^{(s)l_1l_2}_{l_3l_4}(\bm{k},\bm{k}')=\left(\frac{\tilde{U}^{(c)}(\bm{k}-\bm{k}')+\tilde{U}^{(s)}}{4}\right)^{l_1l_3}_{l_2l_4}+\nonumber\\
&\qquad\qquad\qquad\left(\frac{\tilde{U}^{(c)}(\bm{k}+\bm{k}')+\tilde{U}^{(s)}}{4}\right)^{l_1l_4}_{l_2l_3}+\nonumber\\
&\ \ \frac{1}{4}\left[3\tilde{U}^{(s)}\chi^{(s)}\left(\bm{k}-\bm{k}'\right)\tilde{U}^{(s)}-\tilde{U}^{(c)}
\chi^{(c)}\left(\bm{k}-\bm{k}'\right)\tilde{U}^{(c)}\right]^{l_1l_3}_{l_2l_4}+\nonumber\\
&\ \ \frac{1}{4}\left[3\tilde{U}^{(s)}\chi^{(s)}\left(\bm{k}+\bm{k}'\right)\tilde{U}^{(s)}-\tilde{U}^{(c)}
\chi^{(c)}\left(\bm{k}+\bm{k}'\right)\tilde{U}^{(c)}\right]^{l_1l_4}_{l_2l_3},
\end{align}
for the singlet channel and is,
\begin{align}
&\Gamma^{(t)l_1l_2}_{l_3l_4}(\bm{k},\bm{k}')=\left(\frac{\tilde{U}^{(c)}(\bm{k}-\bm{k}')+\tilde{U}^{(s)}}{4}\right)^{l_1l_3}_{l_2l_4}-\nonumber\\
&\qquad\qquad\qquad\left(\frac{\tilde{U}^{(c)}(\bm{k}+\bm{k}')+\tilde{U}^{(s)}}{4}\right)^{l_1l_4}_{l_2l_3}+\nonumber\\
&\ \ \frac{1}{4}\left[\tilde{U}^{(s)}\chi^{(s)}\left(\bm{k}-\bm{k}'\right)\tilde{U}^{(s)}+\tilde{U}^{(c)}
\chi^{(c)}\left(\bm{k}-\bm{k}'\right)\tilde{U}^{(c)}\right]^{l_1l_3}_{l_2l_4}+\nonumber\\
&\ \ \frac{1}{4}\left[\tilde{U}^{(s)}\chi^{(s)}\left(\bm{k}+\bm{k}'\right)\tilde{U}^{(s)}+\tilde{U}^{(c)}
\chi^{(c)}\left(\bm{k}+\bm{k}'\right)\tilde{U}^{(c)}\right]^{l_1l_4}_{l_2l_3},
\end{align}
for the triplet channel.

Note that the vertex $\Gamma^{l_1l_2}_{l_3l_4}(\bm{k},\bm{k}')$ has been symmetrized and anti-symmetrized for the singlet and triplet cases, respectively. The vertex $\Gamma^{l_1l_2}_{l_3l_4}(\bm{k},\bm{k}')$ gives the effective paring interaction vertex $V^{\alpha\beta}(\bm{k,k'})$.

\section{The G-L theory for the chiral SO(4) DW}
\label{Ginzburg}

\subsection{Mixing between VDW and SDW and SO(4) DW}
\label{SO4}

In the main text we prove the degeneracy between the VDW and the SDW. Due to this degeneracy, the two DW order parameters will generally be mixed to lower the energy. Below we study how they would be mixed. The MF Hamiltonian involving both orders should be
\begin{align}\label{HMF0}
\hat H_{\rm MF-DW}&=\hat H_{\rm TB} +\sum_{\iota_1 \iota_2 \bm{k}\sigma\sigma'}\left(\Delta^{(v)}\delta_{\sigma\sigma'}+e^{i\theta}\bm{\Delta}^{(s)}\cdot\bm{\sigma}_{\sigma\sigma'}\right)\nonumber\\
&\qquad \times c^{\dagger}_{\iota_1 \bf{K}\bm{k}\sigma}\xi_{\iota_1 \bf{K} \iota_2 \bf{K'}}(\bm{Q})
\hat c_{\iota_2 \bf{K'}\bm{k}-\bm{Q} \sigma'}+{\rm h.c.},
\end{align}
where $\theta$ is the mixing angle. We shall show below via combined Ginzburg-Landau (G-L) theory and the microscopic calculations that the mixing angle $\theta=\pi/2$, under which the SU(2)$_K\times$SU(2)$_{K^\prime}$ symmetry would be embodied as the SO(4) symmetry for the DW order parameters.

Firstly, due to the global spin-SU(2) symmetry, we can only consider the case in which the spin polarization direction is along the $z$-axis, under which we have
\begin{equation}
\bm{\Delta}^{(s)}\cdot\bm{\sigma}_{\sigma\sigma'}\to \Delta^{(s)}\sigma^z_{\sigma\sigma'}.
\end{equation}

Secondly, the global valley U(1) symmetry of the system requires that when the MF Hamiltonian on the above is operated by the valley U(1) transformation,
\begin{equation}
c^{\dagger}_{\iota v\bm{k}\sigma}\to e^{i v \alpha}c^{\dagger}_{\iota v\bm{k}\sigma},
\end{equation}
the G-L free energy would be unchanged.  Under this transformation, we get effectively that
\begin{equation}
\xi_{\iota_1 \bf{K} \iota_2 \bf{K'}} \to e^{i 2 \alpha}\xi_{\iota_1 \bf{K} \iota_2 \bf{K'}}.
\end{equation}
Note that we can combine the phase factor $e^{i 2 \alpha}$ into the definition of $\Delta^{(v)}$ and $\Delta^{(s)}$ to change the two DW order parameters as complex numbers. Therefore, the G-L free energy of the system can be defined as
\begin{equation}
F(\tilde\Delta^{(v)}, \tilde\Delta^{(s)})\equiv F(e^{i 2 \alpha}\Delta^{(v)}, e^{i 2 \alpha}\Delta^{(s)}).
\end{equation}
The valley U(1) symmetry guarantees that this G-L free energy should be invariant under $\tilde\Delta^{(v/s)} \to e^{i \alpha}\tilde\Delta^{(v/s)}$, which suggests that $\tilde\Delta^{(s/v)}$ should come in pair with $\tilde\Delta^{(s/v)*}$, i.e.
\begin{equation}\label{FSC}
F(\tilde\Delta^{(v)}, \tilde\Delta^{(s)})=F(\left|\tilde\Delta^{(v)}\right|^2,\tilde\Delta^{(v)*}\tilde\Delta^{(s)},\tilde\Delta^{(v)}\tilde\Delta^{(s)*}, \left|\tilde\Delta^{(s)}\right|^2).
\end{equation}

Thirdly, under the SU(2)$_K\times$SU(2)$_{K^\prime}$ symmetry, let's perform the spin-SU(2) transformation only in the valley K,
\begin{equation}
c^{\dagger}_{\iota_1 \bf{K}\bm{k}\sigma}\to e^{i \alpha\sigma}c^{\dagger}_{\iota_1 \bf{K}\bm{k}\sigma},
\end{equation}
and not in the valley K', we have
\begin{eqnarray}\label{SU2}
\left(\begin{array}{c}\tilde\Delta^{(v)}\\\tilde\Delta^{(s)}\end{array}\right)\to\left(\begin{array}{cc}\cos\alpha, i\sin\alpha\\i\sin\alpha, \cos\alpha\end{array}\right) \left(\begin{array}{c}\tilde\Delta^{(v)}\\\tilde\Delta^{(s)}\end{array}\right)\equiv R(\alpha)\left(\begin{array}{c}\tilde\Delta^{(v)}\\\tilde\Delta^{(s)}\end{array}\right).
\end{eqnarray}
The G-L free energy $F$ should be invariant under this transformation. From Eq. (\ref{FSC}), only such combination as $\tilde\Delta^{\dagger} \left(f\right)\tilde\Delta$ can emerge in $F$, where $\tilde\Delta=\left(\begin{array}{c}\tilde\Delta^{(v)}\\\tilde\Delta^{(s)}\end{array}\right)$, and $\left(f\right)$ are $2\times 2$ complex matrix. Then from the invariance of $F$ under Eq. (\ref{SU2}), we have
\begin{equation}
\left[R(\alpha),\left(f\right)\right]=0
\end{equation}
Note that $R(\alpha)=\cos \alpha I+i\sin \alpha \sigma_x$, we have $\left(f\right)=f_1 I+f_2 \sigma_x$. Therefore, we have
\begin{equation}\label{FSC2}
F(\tilde\Delta^{(v)}, \tilde\Delta^{(s)})=F(\left|\tilde\Delta^{(v)}\right|^2+\left|\tilde\Delta^{(s)}\right|^2, \tilde\Delta^{(v)*}\tilde\Delta^{(s)}+\tilde\Delta^{(v)}\tilde\Delta^{(s)*}).
\end{equation}

Fourthly, if we perform the global SU(2) rotation: rotate about the spin-x axis by the angle $\pi$, under which
\begin{eqnarray}\label{globalSU2}
\left(\begin{array}{c}\tilde\Delta^{(v)}\\\tilde\Delta^{(s)}\end{array}\right)\to\left(\begin{array}{c}\tilde\Delta^{(v)}\\-\tilde\Delta^{(s)}\end{array}\right),
\end{eqnarray}
the free energy function $F$ should be invariant. Therefore, the $\left(\tilde\Delta^{(v)*}\tilde\Delta^{(s)}+\tilde\Delta^{(v)}\tilde\Delta^{(s)*}\right)$ term in Eq. (\ref{FSC2}) should only appear in even powers.

From the above analysis, the G-L free energy function $F$ should take the following form up to the fourth order of $\Delta$,
\begin{eqnarray}\label{GLF1}
F(\tilde\Delta^{(v)}, \tilde\Delta^{(s)})=&&-a\left(\left|\tilde\Delta^{(v)}\right|^2+\left|\tilde\Delta^{(s)}\right|^2\right)+b\left(\left|\tilde\Delta^{(v)}\right|^2+\left|\tilde\Delta^{(s)}\right|^2\right)^2\nonumber\\
&&+\gamma\left[\tilde\Delta^{(v)*}\tilde\Delta^{(s)}+\tilde\Delta^{(v)}\tilde\Delta^{(s)*}\right]^2+O(\tilde\Delta^6),
\end{eqnarray}
where $a,b,\gamma$ are real numbers. In the case of $\gamma>0$, the minimization of $F$ requires $\tilde\Delta^{(v)*}\tilde\Delta^{(s)}+\tilde\Delta^{(v)}\tilde\Delta^{(s)*}=0$, which dictates that the phase angle of $\tilde\Delta^{(v)}$ should be different from that of  $\tilde\Delta^{(s)}$ by $\pi/2$. In the case of $\gamma<0$, they should have the same phase angle or be different by a negative sign. Our microscopic calculations always suggest that the former case is realized for the parameters of realistic material. Therefore our combined G-L theory and microscopic calculations suggest that the MF Hamiltonian for the DW ordered state should take the form of
\begin{align}\label{HMF1}
\hat H_{\rm MF-DW}&=\hat H_{\rm TB} +\sum_{\iota_1 \iota_2 \bm{k}\sigma\sigma'}\left(\Delta^{(v)}\delta_{\sigma\sigma'}+i\bm{\Delta}^{(s)}\cdot\bm{\sigma}_{\sigma\sigma'}\right)\nonumber\\
&\qquad \times c^{\dagger}_{\iota_1 \bf{K}\bm{k}\sigma}\xi_{\iota_1 \bf{K} \iota_2 \bf{K'}}(\bm{Q})
\hat c_{\iota_2 \bf{K'}\bm{k}-\bm{Q} \sigma'}+{\rm h.c.}.
\end{align}

Below we prove that the SU(2)$_K\times$SU(2)$_{K^\prime}$ symmetry acted on the $\hat c$ operator of the system is equivalent to the SO(4) symmetry in the 4D DW order-parameter space when the DW MF Hamiltonian takes the above form Eq. (\ref{HMF1}). Note that $\Delta^{(v)}\delta _{\sigma \sigma'}+i\bm{\Delta}^{(s)}\cdot\bm{\sigma}_{\sigma \sigma'} =\left(\Delta^{(v)}I+i\bm{\Delta}^{(s)}\bm{\sigma}\right)_{\sigma \sigma'}\equiv |\bm \varDelta|{\bm M}_{\sigma \sigma'}$, where
\begin{align}
|\bm \varDelta|&=\sqrt{{\Delta^{(v)}}^2+{\Delta^{(s)}_{ x}}^2+{\Delta^{(s)}_{ y}}^2+{\Delta^{(s)}_{ z}}^2}, \label{D2}\\
{\bm M}&=\frac{1}{|\bm \varDelta|}
\left(\begin{array}{cc}
\Delta^{(v)}+i\Delta^{(s)}_{ z},& i\Delta^{(s)}_{ x}+\Delta^{(s)}_{ y}\\
i\Delta^{(s)}_{ x} -\Delta^{(s)}_{ y},& \Delta^{(v)}-i\Delta^{(s)}_{ z}
\end{array}\right).\label{D3}
\end{align}
It's important that ${\bm M}$ is an SU(2) matrix because
\begin{align}\label{D4}
{\bm M}^{\dagger}{\bm M}=I, \qquad
{\rm Det}({\bm M})=1.
\end{align}
Mathematically, \sout{it's}{\color{red}it is} known that any SU(2) matrix can always be parametrized in the form of the equation \eqref{D3}.

On the one hand, let's perform any $U\in \rm{SU(2)}$ on the spin of the $\hat c_{\iota_1+,\bm{k}\sigma}$ operator ($K$ valley) and $V\in \rm{SU(2)}$ on that of the $\hat c_{\iota_2-,\bm{k}-\bm{Q} \sigma'}$ operator ($K^\prime$ valley), resulting in $|\bm \varDelta|{\bm M}\rightarrow |\bm \varDelta|{\bm M}^{\prime}$ with
\begin{align}\label{D5}
{\bm M}^{\prime}=U^{\dagger}{\bm M} V.
\end{align}
Since $U,V\in$ SU(2), from Eq.~\eqref{D4} we can obtain:
\begin{align}\label{D6}
{\bm M^\prime}^{\dagger}{\bm M^\prime}=I, \qquad
{\rm Det}({\bm M^\prime})=1.
\end{align}
The Eq. \eqref{D6} suggests that $\bm M^{\prime}$ is also an SU(2) matrix, which can also be parametrized in the form of Eq.~\eqref{D3} with only $\Delta^{(v)}\rightarrow \Delta^{(c)\prime}$ and $\bm{\Delta}^{(s)}\rightarrow\bm{\Delta}^{(s)\prime}$. This leads to an SO(4) rotation on the 4-component DW order parameter $\bm\varDelta\equiv\left(\Delta^{(v)}, \bm{\Delta}^{(s)}\right)$. Therefore, we have proved that any SU(2)$_K\times$SU(2)$_{K^\prime}$ operation acted on the $\hat c$ operators will lead to an SO(4) rotation on the 4-component DW order parameter $\bm\varDelta$.

On the other hand, suppose that the 4-component DW order parameter $\left(\Delta^{(v)}, \bm{\Delta}^{(s)}\right)$ in Eq.~\eqref{HMF1} is operated by an SO(4) rotation with $\bm M\rightarrow \bm M^\prime$, we can always choose
\begin{align}\label{D7}
\left.\begin{matrix}
U^{\dagger}=\bm M^{-1}\\
V=\bm M^{\prime}
\end{matrix}\right\}\Rightarrow \bm M^{\prime}=U^{\dagger}\bm M V.
\end{align}
This means that any SO(4) rotation on the  4-component spin-valley DW order parameter $\bm\varDelta$ can be realized by the physical SU(2)$_K\times$SU(2)$_{K^\prime}$ operation acted on the $\hat c$ operators.

Combining the above two hands, we arrive at our proof.

\subsection{Mixing between degenerate wave vectors}
\label{mixing_3Q}
On the above subsection, we studied how the SDW and VDW are mixed in the presence of only one wave vector. As a result, they are found to be mixed as 1:i, leading to the SO(4) spin-valley DW. In this subsection, we shall study how the SO(4) DWs with three degenerate wave vectors are mixed via the G-L theory.

In the presence of three degenerate DW orders, the MF Hamiltonian reads,
\begin{align}\label{HMF_chiral}
\hat H_{\rm MF-DW}&=\hat H_{\rm TB} +\sum_{\alpha=1}^3\sum_{l_1l_2\bm{k}\sigma\sigma'}\left(\Delta_{\alpha}^{(v)}\delta_{\sigma\sigma'}+i\bm{\Delta}_{\alpha}^{(s)}\cdot\bm{\sigma}_{\sigma\sigma'}\right)\nonumber\\
&\qquad \times c^{\dagger}_{l_1\bm{k}\sigma}\xi_{l_1l_2}(\bm{Q}_\alpha)
\hat c_{l_2\bm{k}-\bm{Q}_\alpha \sigma'}+{\rm h.c.}\nonumber\\
& = \hat H_{\rm TB} + \sum_{\alpha=1}^3\sum_{l_1l_2\bm{k}\sigma\sigma'}\left[\left(\bm{\varDelta}_\alpha\cdot \bm\varSigma\right)_{\sigma\sigma'}\right. \nonumber\\
&\qquad \times \left.\hat c^{\dagger}_{l_1\bm{k}\sigma}\xi_{l_1l_2}(\bm{Q}_\alpha)
\hat c_{l_2\bm{k}-\bm{Q}_\alpha \sigma'}+{\rm h.c.}\right],
\end{align}
where the 4-component vector $\bm{\varDelta}_{\alpha}\equiv\left(\Delta^{(v)}_{\alpha}, \bm\Delta^{(s)}_{\alpha}\right)=\left(\Delta^{(v)}_{\alpha}, \Delta^{(s)}_{\alpha,x},\Delta^{(s)}_{\alpha,y},\Delta^{(s)}_{\alpha,z}\right)\in\mathbb{R}^4$ and $\bm\varSigma=\left(\sigma^{(0)},i\bm\sigma\right)$ with $\sigma^{(0)}$ to be the $2\times2$ identity matrix.

The G-L free energy $F$ as a function of the three 4-component vectors $\bm{\varDelta}_{\alpha}$ can be expanded up to the quartic order of the Taylor's series as
\begin{eqnarray}
F&=&F\left(\bm{\varDelta}_{1},\bm{\varDelta}_{2},\bm{\varDelta}_{3}\right)\nonumber\\
&=&F_2+F_4,
\end{eqnarray}
where $F_2$ and $F_4$ are quadratic and quartic order terms respectively.

Firstly, due to the SO(4) symmetry of the system, the $F_2$ and $F_4$ can only contain such terms as $\left|\bm{\varDelta}_{\alpha}\right|^2$ and $\bm{\varDelta}_{\alpha}\cdot \bm{\varDelta}_{\beta}$. As a result, we have
\begin{eqnarray}\label{F2_DW}
F_2=&&\alpha \left|\bm{\varDelta}_{1}\right|^2 +\beta \left|\bm{\varDelta}_{2}\right|^2 + \gamma \left|\bm{\varDelta}_{3}\right|^2 \nonumber\\&&+\theta \bm{\varDelta}_{1}\cdot \bm{\varDelta}_{2} + \delta \bm{\varDelta}_{1}\cdot \bm{\varDelta}_{3} + \xi\bm{\varDelta}_{2}\cdot \bm{\varDelta}_{3}
\end{eqnarray}
From the $D_3$ rotation symmetry of the system, we know that
\begin{equation}
\alpha=\beta=\gamma,\theta=\delta=\xi.
\end{equation}

Let's then investigate the consequence of the unit-cell translation symmetry of the system. Setting the unit vector of the original honeycomb lattice as $\mathbf{a}_1, \mathbf{a}_2$, and the corresponding unit vector in the reciprocal lattice as $\mathbf{b}_1, \mathbf{b}_2$, we have $\mathbf{Q}_1=\frac{\mathbf{b}_1}{2}$, $\mathbf{Q}_2=\frac{\mathbf{b}_2}{2}$ and $\mathbf{Q}_3=\frac{\mathbf{b}_1+\mathbf{b}_2}{2}$. Let's translate the system by the unit vector $\mathbf{a}_i (i=1,2)$, under which we have
\begin{eqnarray}
c^{\dagger}_{l_1\bm{k}\sigma}\to && e^{-i\bm{k}\cdot \mathbf{a}_i}c^{\dagger}_{l_1\bm{k}\sigma}, \nonumber\\
c_{l_2\bm{k}-\bm{Q}_\alpha \sigma'} && e^{i\left(\bm{k}-\bm{Q}_\alpha\right)\cdot \mathbf{a}_i}c_{l_2\bm{k}-\bm{Q}_\alpha \sigma'}.
\end{eqnarray}
Then Eq. (\ref{HMF_chiral}) suggests that under the unit vector $\mathbf{a}_i$ translation, we effectively have
\begin{equation}
\bm{\varDelta}_{\alpha} \to e^{-i\bm{Q}_\alpha\cdot \mathbf{a}_i} \bm{\varDelta}_{\alpha}.
\end{equation}
For the case $i=1$, we have
\begin{eqnarray}
\bm{\varDelta}_{1} \to &&-\bm{\varDelta}_{1}\nonumber\\
\bm{\varDelta}_{2} \to &&\bm{\varDelta}_{2}\nonumber\\
\bm{\varDelta}_{3} \to &&-\bm{\varDelta}_{3}.
\end{eqnarray}
Then the $\mathbf{a}_1$ translational invariance of $F_2$ in Eq.(\ref{F2_DW}) dictates
 \begin{equation}
\theta=\xi=0.
\end{equation}
Similarly, the $\mathbf{a}_2$ translational invariance of $F_2$ dictates
 \begin{equation}
\theta=\delta=0.
\end{equation}
Therefore, up to the quadratic order term of $\{\bm{\varDelta}_{\alpha}\}$, we have
 \begin{equation}\label{F2}
F_2=\alpha\left(\left|\bm{\varDelta}_{1}\right|^2+\left|\bm{\varDelta}_{2}\right|^2+\left|\bm{\varDelta}_{3}\right|^2\right)
\end{equation}

Through similar analysis on symmetry as the above, we can obtain the following symmetry-allowed form of $F_4$,
 \begin{eqnarray}\label{F4}
F_4&=&a\left(\left|\bm{\varDelta}_{1}\right|^4+\left|\bm{\varDelta}_{2}\right|^4+\left|\bm{\varDelta}_{3}\right|^4\right)\nonumber\\
&+&b\left(\left|\bm{\varDelta}_{1}\right|^2\left|\bm{\varDelta}_{2}\right|^2+
\left|\bm{\varDelta}_{1}\right|^2\left|\bm{\varDelta}_{3}\right|^2+\left|\bm{\varDelta}_{2}\right|^2\left|\bm{\varDelta}_{3}\right|^2\right)\nonumber\\
&+&c\left[\left(\bm{\varDelta}_{1}\cdot\bm{\varDelta}_{2}\right)^2+\left(\bm{\varDelta}_{1}\cdot\bm{\varDelta}_{3}\right)^2+\left(\bm{\varDelta}_{2}\cdot\bm{\varDelta}_{3}\right)^2\right].
\end{eqnarray}
From combined Eq. (\ref{F2}) and Eq. (\ref{F4}), we have
\begin{eqnarray}\label{F24}
F&=&F_2+F_4\nonumber\\
&=&\alpha\left(\left|\bm{\varDelta}_{1}\right|^2+\left|\bm{\varDelta}_{2}\right|^2+\left|\bm{\varDelta}_{3}\right|^2\right)\nonumber\\
&+&a\left(\left|\bm{\varDelta}_{1}\right|^4+\left|\bm{\varDelta}_{2}\right|^4+\left|\bm{\varDelta}_{3}\right|^4\right)\nonumber\\
&+&b\left(\left|\bm{\varDelta}_{1}\right|^2\left|\bm{\varDelta}_{2}\right|^2+
\left|\bm{\varDelta}_{1}\right|^2\left|\bm{\varDelta}_{3}\right|^2+\left|\bm{\varDelta}_{2}\right|^2\left|\bm{\varDelta}_{3}\right|^2\right)\nonumber\\
&+&c\left[\left(\bm{\varDelta}_{1}\cdot\bm{\varDelta}_{2}\right)^2+\left(\bm{\varDelta}_{1}\cdot\bm{\varDelta}_{3}\right)^2+\left(\bm{\varDelta}_{2}\cdot\bm{\varDelta}_{3}\right)^2\right]\nonumber\\
&+& O(\bm{\varDelta}^6).
\end{eqnarray}

In the following, we shall solve the configurations of $\left\{\bm{\varDelta}_{\alpha}\right\}$ which minimize the free energy function $F$ provided by Eq. (\ref{F24}).

Firstly, we study the relative orientations among the three DW order parameters $\bm{\varDelta}_{\alpha} (\alpha=1,2,3)$. This problem is simply determined by the sign of $c$ in Eq. (\ref{F24}). The answer is as following,
\begin{eqnarray}\label{sign_of_c}
c>0 &\Rightarrow&  \bm{\varDelta}_{1}\perp  \bm{\varDelta}_{2} \perp  \bm{\varDelta}_{3}, \nonumber\\
c<0 &\Rightarrow&  \bm{\varDelta}_{1}\varparallel  \bm{\varDelta}_{2} \varparallel  \bm{\varDelta}_{3}.
\end{eqnarray}

Then, we study the relative amplitudes among $\bm{\varDelta}_{\alpha} (\alpha=1,2,3)$. Let their amplitudes be $\Delta_{\alpha} (\alpha=1,2,3)$. Eq. (\ref{sign_of_c}) yields that in both cases of $c>0$ and $c<0$, we have
\begin{eqnarray}\label{F_amplitude}
F&=&F_2+F_4\nonumber\\
&=&\alpha\left(\Delta_{1}^2+\Delta_{2}^2+\Delta_{3}^2\right)+a\left(\Delta_{1}^4+\Delta_{2}^4+\Delta_{3}^4\right)\nonumber\\
&+&\tilde b\left(\Delta_{1}^2\Delta_{2}^2+\Delta_{1}^2\Delta_{3}^2+\Delta_{2}^2\Delta_{3}^2\right)+ O(\Delta^6),
\end{eqnarray}
where
\begin{eqnarray}\label{tilde_b}
\tilde b=\left\{\begin{array}{cc}b,c>0\\b+c,c<0\end{array}\right..
\end{eqnarray}
In the long-ranged DW ordered state, we should have
\begin{eqnarray}\label{alpha_a}
\alpha<0,a>0.
\end{eqnarray}

To solve the minimum of the free-energy function $F$ provided by Eq. (\ref{F_amplitude}), we deform it as
\begin{eqnarray}\label{F_amplitude_2}
F&=&F_2+F_4\nonumber\\
&=&\alpha\left(\Delta_{1}^2+\Delta_{2}^2+\Delta_{3}^2\right)+a\left(\Delta_{1}^2+\Delta_{2}^2+\Delta_{3}^2\right)^2\nonumber\\
&+&\left(\tilde b-2a\right)\left(\Delta_{1}^2\Delta_{2}^2+\Delta_{1}^2\Delta_{3}^2+\Delta_{2}^2\Delta_{3}^2\right)+ O(\Delta^6).
\end{eqnarray}
Clearly, the solution for the minimization of $F$ function on the above is determined by the sign of $\tilde b-2a$. The result is as following,
\begin{eqnarray}\label{b2a1}
\tilde b>2a &\Rightarrow&  \Delta_{1}\ne0, \text{or}, \Delta_{2}\ne0, \text{or}, \Delta_{3}\ne0,\nonumber\\
\tilde b<2a &\Rightarrow&  \Delta_{1}=\Delta_{2}=\Delta_{3}.
\end{eqnarray}
Note that on the above Eq. (\ref{b2a1}), in the first case $\tilde b>2a$, only one of $\Delta_{\alpha} (\alpha=1,2,3)$ can be nonzero. Such a state is the nematic state, which only hosts one wave vector. In the second case, the amplitudes of the three DW orders are equal, which can either be the chiral SO(4) DW in which the orientations of the three DW order parameters are perpendicular to one another or be the collinear SO(4) DW state in which the orientations of the three DW order parameters are parallel to one another, which is determined by Eq. (\ref{sign_of_c}).

Summarizing the above derivations, we get the following possible solutions for the minimization of the free energy function $F$ defined in Eq. (\ref{F24})
\begin{eqnarray}\label{b2a}
c<0&,&\left\{\begin{array} {c}b<2a-c\Rightarrow \text{collinear-DW}\\b>2a-c\Rightarrow \text{nematic-DW}\end{array}\right.\nonumber\\
c>0&,&\left\{\begin{array} {c}b<2a\Rightarrow \text{chiral-DW}\\b>2a\Rightarrow \text{nematic-DW}\end{array}\right.
\end{eqnarray}
Therefore, only three possible solutions exist, i.e. the collinear SO(4) spin-valley DW state, the chiral SO(4) spin-valley DW state and the nematic SO(4) spin-valley DW state. In the collinear state, the three DW order parameters $\bm{\varDelta}_{1}=
\bm{\varDelta}_{2}=\bm{\varDelta}_{3}$. In the chiral state, they satisfy $\bm{\varDelta}_{1}\perp  \bm{\varDelta}_{2} \perp  \bm{\varDelta}_{3}$ and $|\bm{\varDelta}_{1}|=  |\bm{\varDelta}_{2}|=|\bm{\varDelta}_{3}|$. In the nematic state, only one of the three $\bm{\varDelta}_{\alpha}(\alpha=1, 2, 3)$ exists, and the other two vanish. In realistic system, which state would be the ground state cannot be know only from the G-L theory. Instead, the microscopic calculations are needed.

\subsection{The case of pure SDW or VDW}
\label{puresdwvdw}
In some case in our study we only consider the pure SDW or VDW order parameters, particularly in the case when we try to compare the energies of a pure SDW state and a pure VDW state.

In the case when we consider the pure SDW state, the SDW-MF Hamiltonian reads,
\begin{align}\label{HMF_SDW}
\hat H_{\rm MF-SDW}&=\hat H_{\rm TB} +\sum_{\alpha=1}^3\sum_{l_1l_2\bm{k}\sigma\sigma'}\left(\bm{\Delta}_{\alpha}^{(s)}\cdot\bm{\sigma}\right)_{\sigma\sigma'}\nonumber\\
&\qquad \times c^{\dagger}_{l_1\bm{k}\sigma}\xi_{l_1l_2}(\bm{Q}_\alpha)
\hat c_{l_2\bm{k}-\bm{Q}_\alpha \sigma'}+{\rm h.c.}.
\end{align}
Here $\bm{\Delta}_{\alpha}^{(s)}$ denotes the three-component SDW order parameters, which are abbreviated as $\bm{\Delta}_{\alpha}$ below.
The G-L free energy should be
\begin{eqnarray}
F&=&F\left(\bm{\Delta}_{1},\bm{\Delta}_{2},\bm{\Delta}_{3}\right)=F_2+F_4.
\end{eqnarray}

Adopting the symmetry-based analysis parallel to that performed on the above subsection, we can obtain
\begin{eqnarray}\label{F24_SDW}
F&=&F_2+F_4\nonumber\\
&=&\alpha\left(\left|\bm{\Delta}_{1}\right|^2+\left|\bm{\Delta}_{2}\right|^2+\left|\bm{\Delta}_{3}\right|^2\right)+a\left(\left|\bm{\Delta}_{1}\right|^4+\left|\bm{\Delta}_{2}\right|^4+\left|\bm{\Delta}_{3}\right|^4\right)\nonumber\\
&+&b\left(\left|\bm{\Delta}_{1}\right|^2\left|\bm{\Delta}_{2}\right|^2+
\left|\bm{\Delta}_{1}\right|^2\left|\bm{\Delta}_{3}\right|^2+\left|\bm{\Delta}_{2}\right|^2\left|\bm{\Delta}_{3}\right|^2\right)\nonumber\\
&+&c\left[\left(\bm{\Delta}_{1}\cdot\bm{\Delta}_{2}\right)^2+\left(\bm{\Delta}_{1}\cdot\bm{\Delta}_{3}\right)^2+\left(\bm{\Delta}_{2}\cdot\bm{\Delta}_{3}\right)^2\right]\nonumber\\
&+& O(\bm{\Delta}^6).
\end{eqnarray}
The solution for the minimum of Eq. (\ref{F24_SDW}) yields
\begin{eqnarray}\label{b2a}
c<0&,&\left\{\begin{array} {c}b<2a-c\Rightarrow \text{collinear-SDW}\\b>2a-c\Rightarrow \text{nematic-SDW}\end{array}\right.\nonumber\\
c>0&,&\left\{\begin{array} {c}b<2a\Rightarrow \text{chiral-SDW}\\b>2a\Rightarrow \text{nematic-SDW}\end{array}\right.
\end{eqnarray}
Therefore, there are also three possible SDW solutions, i.e. the collinear-SDW state, the chiral-SDW state and the nematic-SDW state. In realistic system, the microscopic calculations are needed to determine which state should be the ground state.

In the case when we consider the pure VDW state, the VDW-MF Hamiltonian reads,
\begin{align}\label{HMF_VDW}
\hat H_{\rm MF-VDW}&=\hat H_{\rm TB} +\sum_{\alpha=1}^3\sum_{l_1l_2\bm{k}\sigma}\Delta_{\alpha}^{(v)} c^{\dagger}_{l_1\bm{k}\sigma}\xi_{l_1l_2}(\bm{Q}_\alpha)
\hat c_{l_2\bm{k}-\bm{Q}_\alpha \sigma}+{\rm h.c.}.
\end{align}
Here $\Delta_{\alpha}^{(v)}$ denotes the VDW order parameters, which are abbreviated as $\Delta_{\alpha}$ below. Adopting similar symmetry-based analysis parallel to the above, we can obtain
\begin{eqnarray}\label{F24_VDW}
F&=&F_2+F_4\nonumber\\
&=&\alpha\left(\Delta^2_{1}+\Delta^2_{2}+\Delta^2_{3}\right)+a\left(\Delta^4_{1}+\Delta^4_{2}+\Delta^4_{3}\right)\nonumber\\
&+&b\left(\Delta^2_{1}\Delta^2_{2}+
\Delta^2_{1}\Delta^2_{3}+\Delta^2_{2}\Delta^2_{3}\right)+ O(\Delta^6).
\end{eqnarray}
The solution for the minimum of Eq. (\ref{F24_VDW}) yields
\begin{eqnarray}\label{b2aVDW}
b<2a&\Rightarrow& \text{isotropic-VDW}\nonumber\\
b>2a&\Rightarrow& \text{nematic-VDW}
\end{eqnarray}
Therefore, there are two possible VDW solutions, i.e. the isotropic-VDW state and the nematic-VDW state. While the former contains three VDW components with equal amplitudes for the three wave vectors, the latter only contains one for one arbitrarily chosen wave vector. In realistic system, the microscopic calculations are needed to determine which state should be the ground state.

%

\end{document}